\documentclass[3p]{elsart}
\usepackage{graphicx}
\usepackage{multirow} 
\usepackage{amssymb}
\usepackage{amsmath}
\usepackage{enumerate}
\usepackage{color}
\usepackage{ulem}
\usepackage[switch]{lineno}
\usepackage{url}
\graphicspath{{../figures/}}

\begin{document}

\begin{frontmatter}

\title{Simulations of reflected radio signals from cosmic ray induced air showers}
\author[USC]{Jaime Alvarez-Mu\~niz}
\ead{Corresponding author: jaime.alvarezmuniz@gmail.com - Phone: +34 881 813 968 - Fax:+34 881 814 086}
\author[USC]{Washington R. Carvalho Jr.}
\ead{carvajr@gmail.com}
\author[USC]{Daniel Garc\'\i a-Fern\'andez}
\ead{sirzid@gmail.com}
\author[UH]{Harm Schoorlemmer}
\ead{harmscho@phys.hawaii.edu}
\author[USC]{Enrique Zas}
\ead{zas@fpaxp1.usc.es}

\address[USC]{Departamento de F\'\i sica de Part\'\i culas
\& Instituto Galego de F\'\i sica de Altas Enerx\'\i as, \\
Universidade de Santiago de Compostela, 
15782 Santiago de Compostela, Spain}

\address[UH]{University of Hawaii at Manoa, 
Department of Physics and Astronomy, Honolulu,
Hawaii 96822, USA}

\begin{abstract}
We present the calculation of 
coherent radio pulses emitted by extensive air showers induced 
by ultra-high energy cosmic rays accounting for
reflection on the Earth's surface. Results have been obtained with a 
simulation program that calculates the contributions from shower particles
after reflection at a surface plane. The properties of the radiation are 
discussed in detail emphasizing the effects of reflection. 
The shape of the frequency spectrum is shown to be 
closely related to the angle of the observer with respect
to shower axis, becoming hardest in the Cherenkov direction. 
The intensity of the flux at a fixed observation angle is shown 
to scale with the square of the primary particle energy to very 
good accuracy indicating the coherent aspect of the emission.
The simulation methods of this paper provide the foundations 
for energy reconstruction of experiments looking at
the Earth from balloons and satellites. They can also 
be used in dedicated studies of existing and future 
experimental proposals.
\end{abstract}
\end{frontmatter}

\linenumbers

\section{Introduction}

There is a renewed interest in using the radio technique for the
detection of extensive air showers induced by Ultra-High Energy Cosmic
Rays (UHECR). Dedicated experimental initiatives such as CODALEMA~\cite{CODALEMA},
LOFAR~\cite{LOFAR}, AERA~\cite{AERA}, LOPES~\cite{LOPES} and
Tunka-Rex~\cite{Tunka-Rex_ICRC13}, with instrumentation far superior to that used in the early days
of radio detection, have stimulated an impressive progress in the field. 
As a result the possibility of using the radio technique, either as a 
complement to other techniques for the detection of extensive air
showers, or even to design new UHECR detectors based on it, is now
seriously considered~\cite{Huege_ICRC13}. 

The serendipitous discovery of 16 fast radio pulses from air showers 
extending up to the GHz range with the first ANITA balloon born antenna system~\cite{ANITA_CR} 
came as a surprise. This system, operating with a frequency band between 200 and
1200\,MHz, had been originally conceived for detecting
radio emission from neutrino interactions in the Antarctic ice sheet. 
However, the detected emission was strongly polarized 
in the direction defined by the cross product of the geomagnetic field
and the shower direction, which is characteristic of radio signals
from air showers due to the current induced by the Earth's magnetic field~\cite{CODALEMA,GeoEffect}. 
Detailed simulations of these pulses 
showed recently that coherent emission can extend up to and above GHz frequencies \cite{CROME,ZHAireS_ANITA} 
in spite of simple dimensional arguments suggesting the contrary \cite{ZHS_nus}.
This is because the varying refractive index of the atmosphere
introduces a significant effect on the relative travel time 
of emissions originating from different locations. When
these delays are accounted for, a Cherenkov like ring can appear in which
the signals from a large region of the shower arrive with little time delay 
and thus add coherently~\cite{ZHAireS_ANITA}. This angle is mostly determined by the refractive 
index at the shower maximum, simply because that is the region with
more charge and current and hence contributing most.

Fourteen out of the sixteen detected pulses were reconstructed to be 
coming from distinct points on the polar ice cap and showed inverted 
polarity with respect to the other two. This was interpreted as clear
evidence that they were detected 
after reflection on the polar ice cap. 
The reflection of the radio flashes introduces several new aspects to
the calculation of pulses at the detector that have not been previously addressed. 
Naturally reflection implies a reduction of the emission
due to the Fresnel coefficients. 
The relative time delays with respect to detection at ground level are also altered, 
since the pulses propagate upwards after reflection towards the top of the atmosphere 
along a decreasing refractive index profile.
These effects need to be taken into consideration
to interpret the events detected by ANITA and to evaluate the  
acceptance of experiments that rely on observing radiation induced by showers from 
mountain tops, balloon payloads~\cite{ref:Motloch} such as the next 
ANITA flight and EVA~\cite{EVA} or from satellite payloads as proposed in SWORD~\cite{ref:SWORD}. 

In this article we simulate and describe the properties of radio
pulses emitted from extensive air showers after reflection off a surface. Most of the calculations 
are performed in a configuration suited for a high altitude balloon flight over Antarctica,
however the developed methods can be applied to other reflective surfaces
and different detector altitudes. 
We modified the ZHAireS code~\cite{ZHAireS} to calculate 
the radio emission from air showers after reflection on a flat surface. 
We first describe the geometry and explain the assumptions made to 
adapt the program to calculate the reflected radiation. 
After this we generate a set of simulations to investigate the signal properties
as a function of off-axis angle, frequency, zenith angle, and energy of the primary particle, 
and stress the importance of properly accounting for the 
Fresnel coefficients in the reflection, as well as for the propagation of the pulses
towards the top of the atmosphere. 
In the Appendix we use a ray tracing code and a simplified model for the 
emission (that displays the main features of the predicted radiation
as has been justified elsewhere~\cite{ZHAireS_ANITA,1Dmodel}), to 
confirm that the effect of curved light trajectories can be neglected. 

\section{Geometry and scales relevant for reflected events}
\label{section2}

\begin{figure}
\begin{center}
\includegraphics[width=0.8\textwidth] {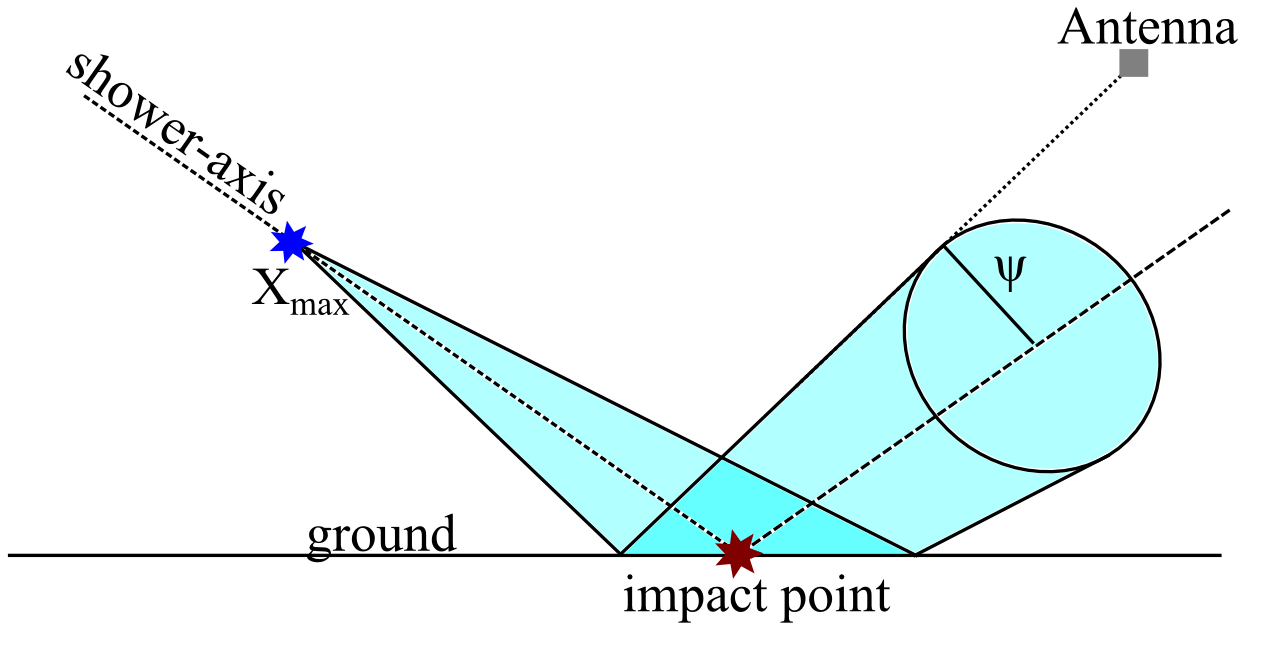}
\caption{To describe a location of an antenna we use the ``off-axis" angle $\psi$, 
as that between the shower axis and the line joining the location of
$X_{\text{max}}$ and the position of the antenna.}
\label{fig:psi}
\end{center}
\end{figure}

The appealing aspect of observing radiation of air showers after reflection 
is that a large atmospheric volume can be monitored with a single detector. 
Therefore, the most interesting geometry is given by air showers that impact
Earth's surface at large zenith angles.
\begin{figure}
\begin{center}
\scalebox{0.7}
{
\includegraphics{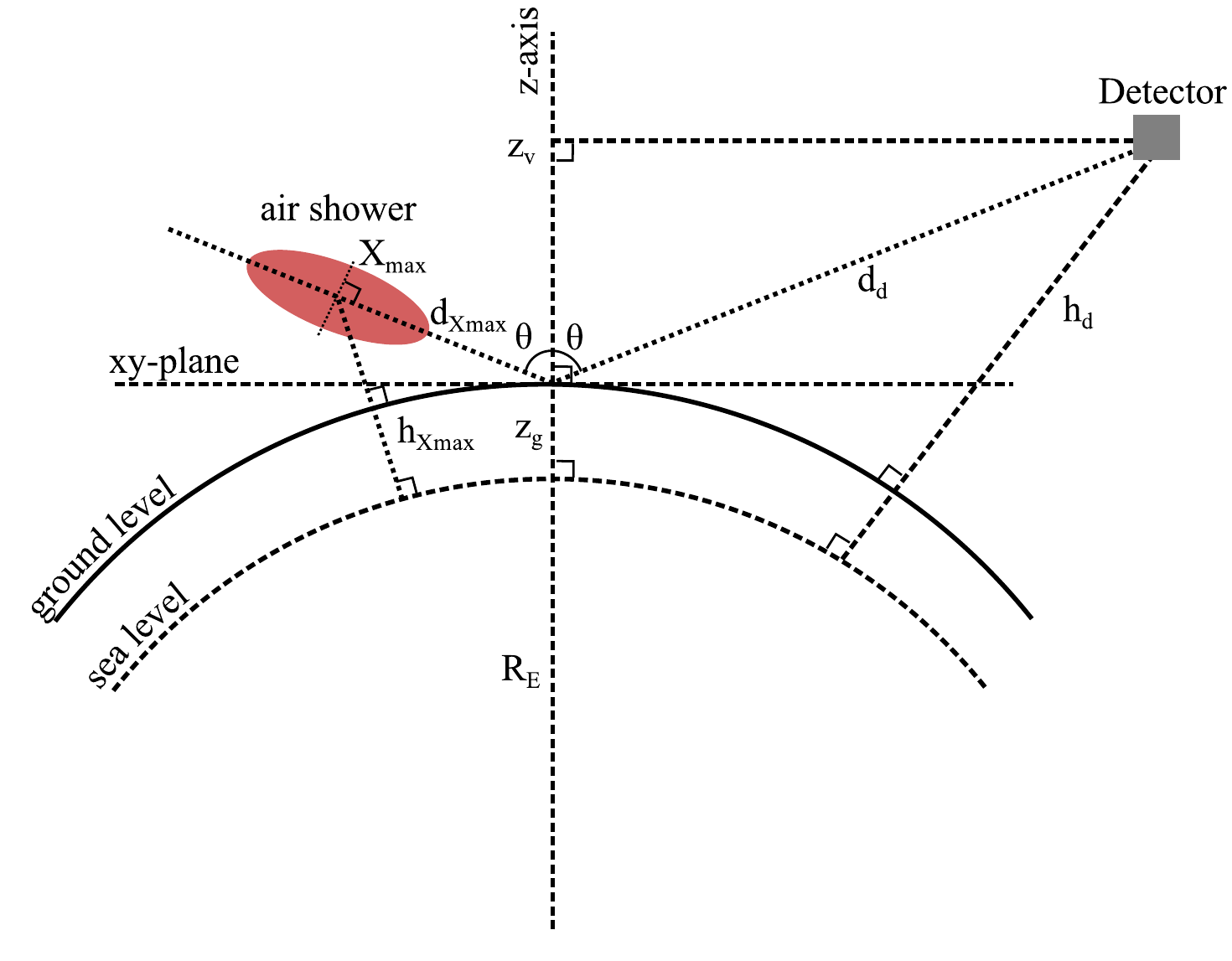}
}
\caption{Basic geometry for reflected signals from air showers (see text for details, see also Fig.~\ref{fig:psi}). }
\label{fig:geometry}
\end{center}
\end{figure}
The basic geometry of the problem is sketched in Fig.~\ref{fig:geometry}. 
We define a rectangular coordinate system with the $z$-axis pointing upwards in
the vertical direction and the $x-y$ plane tangent to the Earth's surface. 
The reflective surface will be approximated by this plane. 
The origin of the coordinate system is the point at which the shower
axis intercepts the Earth's surface which is assumed to be at a ground altitude $z_g$ above sea 
level. The zenith angle of the shower, $\theta$, is defined with respect to the $z-$axis.  
We define the {\sl off-axis angle} $\psi$ in Fig.~\ref{fig:psi}, to describe the
angular deviation of the emitted radiation with respect to the shower axis\footnote{$\psi$ as depicted 
in Fig~\ref{fig:psi} is also used later in this work to refer to the location of the observer.}. 
A generic detector is positioned at a point with vertical altitude
$h_d$. In Fig.~\ref{fig:geometry} the detector is displayed in a special position such that it
sees the reflected radiation which was emitted precisely along the direction of
shower axis with $\psi=0^\circ$. 
The altitude at which shower maximum ($X_{\rm max}$) is reached,
$h_{\rm Xmax}$, is also of relevance. Besides determining the angle at
which the emission is largest \cite{ZHAireS_ANITA}, it also sets the scale of distances the pulse has 
to travel to reach the detector,  $d_{\rm Xmax}+d_d$,  where $d_{\rm  Xmax}$ and 
$d_d$ respectively denote the distances from  
the origin of the coordinate system to shower maximum and to the detector. 

\begin{figure}
\begin{center}
{
\includegraphics[width=0.65\textwidth]{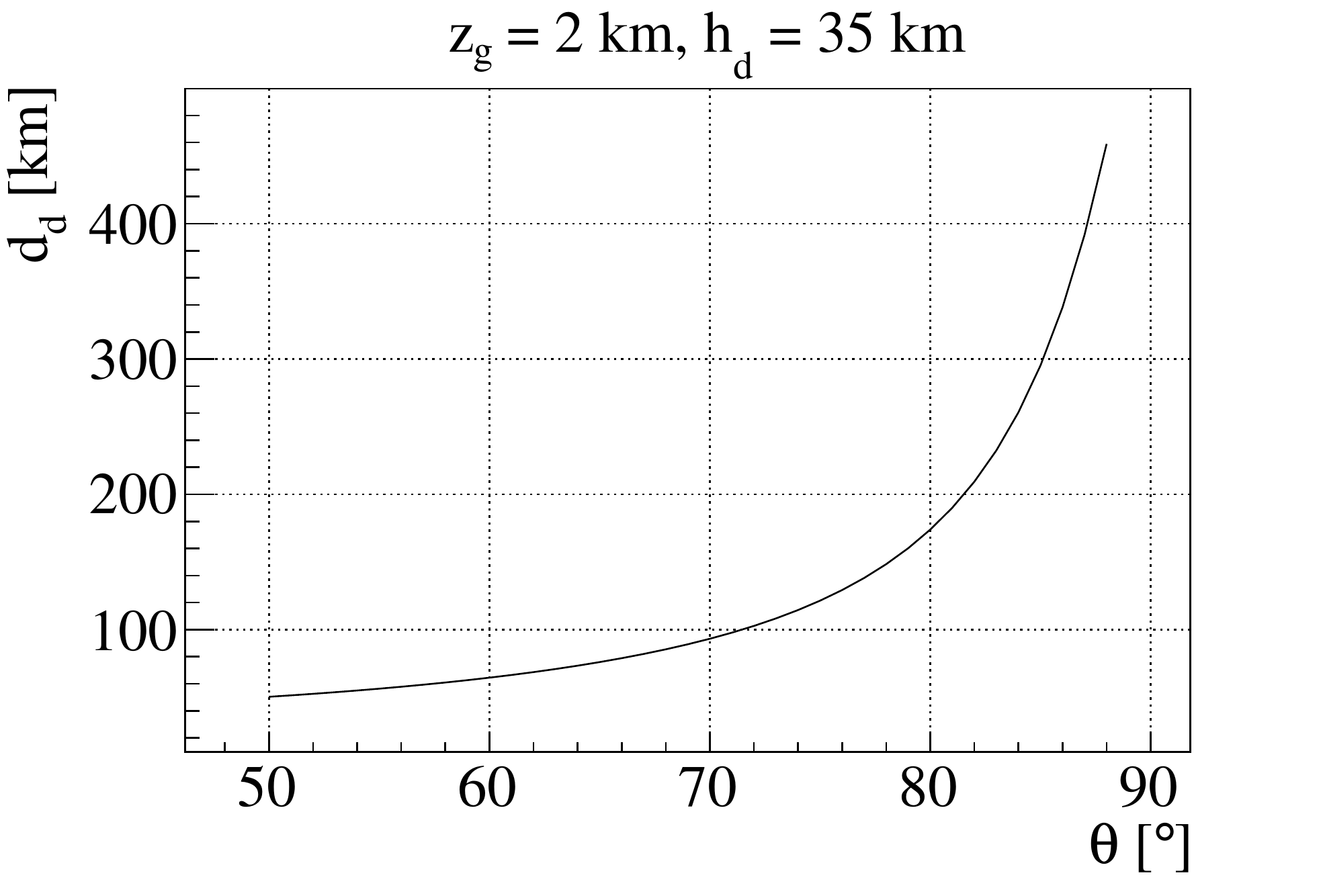} \\
\includegraphics[width=0.65\textwidth]{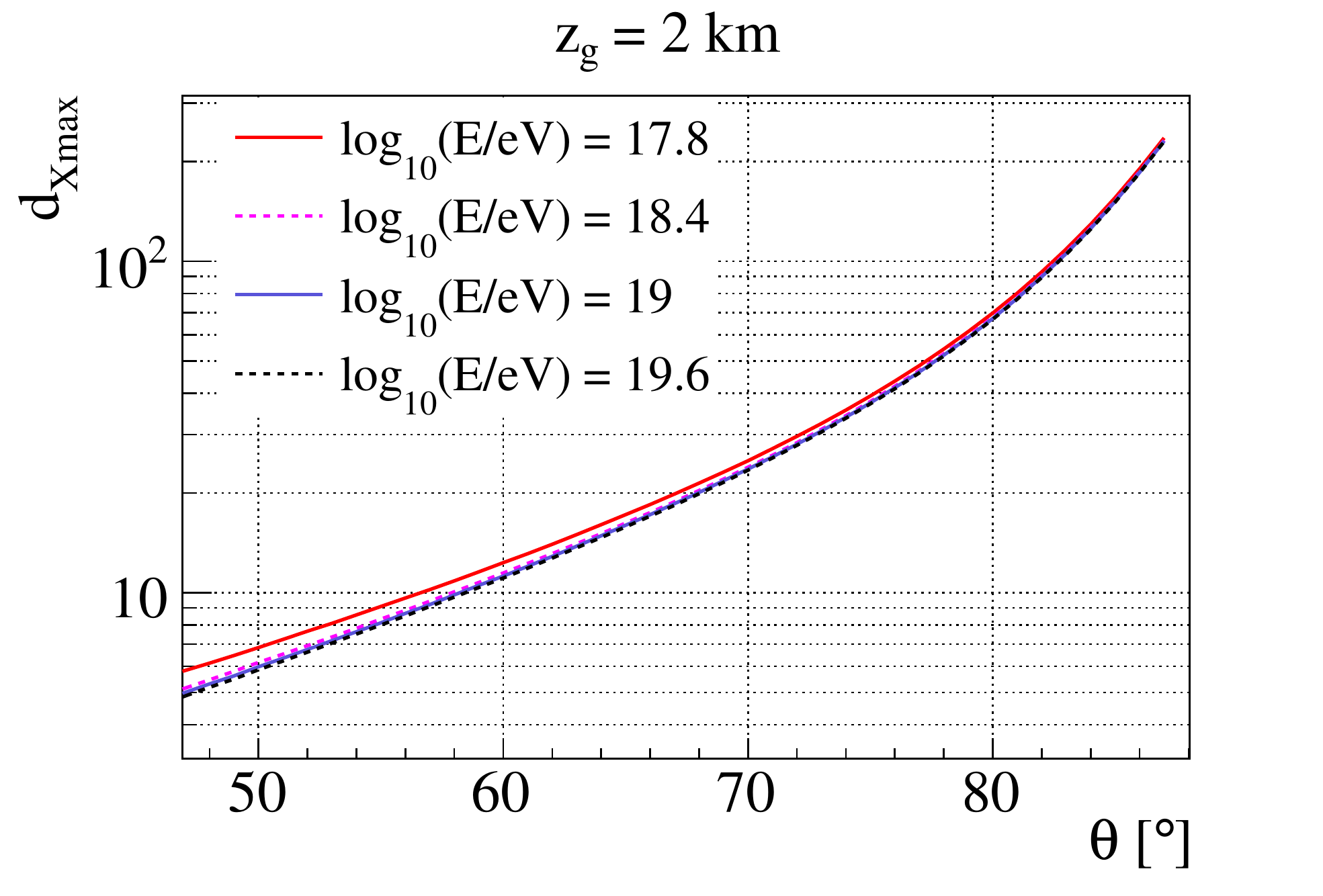} \\
\includegraphics[width=0.65\textwidth]{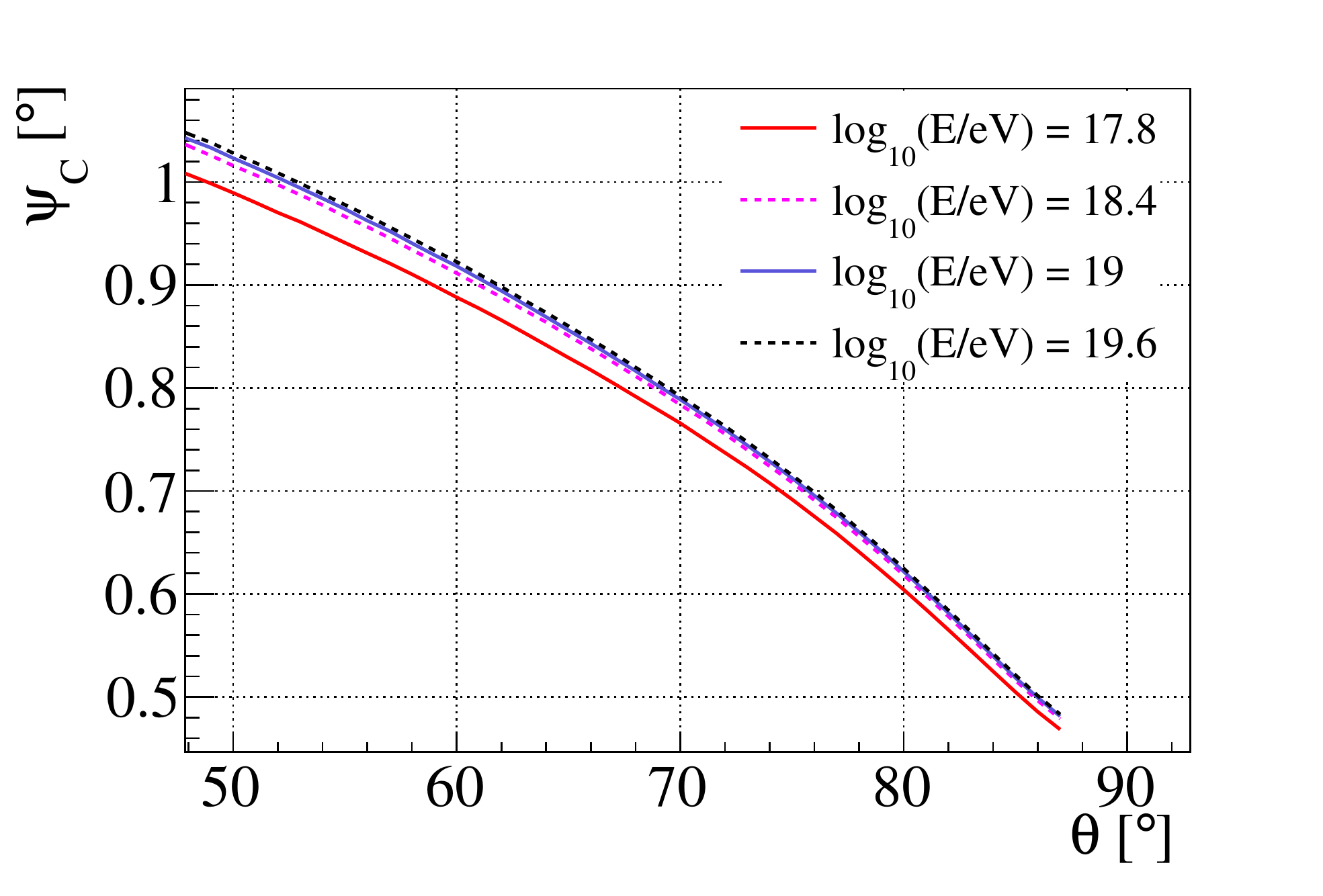} 
}
\caption{
Top panel: distance from the reflection point to the detector ($d_d$ in Fig.~\ref{fig:geometry});
middel panel: distance to shower maximum ($d_{\rm Xmax}$ in Fig.~\ref{fig:geometry}); 
and bottom panel: the Cherenkov angle at $X_{\text{max}}$ as a function of shower zenith angle. 
For the location of $X_{\text{max}}$ 
we used the average value of $\langle X_{\text{max}}\rangle$ as a function of energy 
as measured by the Pierre Auger Collaboration~\cite{AugerXmax,AugerXmax2}.}
\label{fig:dXmax}
\end{center}
\end{figure}

We illustrate the typical scales of this geometry by showing in Fig.~\ref{fig:dXmax} 
some parameters for a high altitude balloon over Antarctica. We take a detector at a
typical altitude of $h_d=35~{\rm km}$, a ground altitude of the ice 
cap of $z_g=2~{\rm km}$ and the Earth's radius $R_E=6357~{\rm km}$. The distance $d_d$ becomes simply a
function of $\theta$ which is illustrated in the top panel of Fig.~\ref{fig:dXmax}. 
To estimate the distance to shower maximum, $d_{\text{Xmax}}$, we use the average slant 
depth of shower maximum $\langle X_{\text{max}}\rangle$ as observed by the Pierre Auger 
Observatory~\cite{AugerXmax,AugerXmax2} together with the atmospheric density profile used in the air 
shower simulation package AIRES \cite{ref:Aires}. Clearly the average
position of $X_{\rm max}$ depends on shower energy, but the effect is accidentally small
according to the measurements at the Pierre Auger Observatory which indicate little change of
$X_{\rm max}$ in the energy range of $10^{17.8}$ to $10^{19.6}$ \cite{AugerXmax,AugerXmax2}. 
The results for different primary energies are shown as a function of $\theta$ in
the middle panel of Fig.~\ref{fig:dXmax}. It should be
noted that the measured RMS fluctuations of $X_{\rm max}$ are between
$20$ and $60$ g~cm$^{-2}$\cite{AugerXmax,AugerXmax2} corresponding to variations
of $d_{\rm Xmax}$ below $11\%$ ($3\%$) for a zenith angle of $60^\circ$ ($85^\circ$) 
degrees. The variation relative to the total distance travelled by the
pulse $d_{\rm Xmax}+d_d$ reduces to $1.5\%$ (1.0\%). 

A relevant parameter is the Cherenkov angle, $\psi_C\approx \cos^{-1}(1/n)$, at the location of $X_{\text{max}}$ that 
is directly obtained using the refractive index $n$ at the corresponding altitude,
$h_{Xmax}$. In the following the refractive index is approximated by a
simple exponential function of altitude $h$ given by 
$n(h) = 1 + \eta_0 \, e^{-\kappa h}$,
with $\eta_0=325\times 10^{-6} $ and 
$\kappa = 0.1218$\,km$^{-1}$~\cite{ZHAireS}. 
The Cherenkov angle at $X_{\text{max}}$ is shown in Fig.~\ref{fig:dXmax} (bottom) for the 
above parameterization. 

\section{Introducing reflection in ZHAireS} 

ZHAireS~\cite{ZHAireS} is a simulation program that has been developed combining
the AIRES package for air shower simulations \cite{ref:Aires} with the 
``ZHS algorithm'' which was originally developed to calculate radio
emission from high energy showers in homogeneous
ice~\cite{ref:Zas1992,ref:Muniz2010} and then extended for use 
in air showers~\cite{ZHAireS,DanielZHS}. 
The contribution from each track element to the radio pulse at any given position and time is 
calculated in ZHAireS assuming the particle travels at a constant speed and in a rectilinear
motion and also accounting for the travel times taken by the radiation to reach from each of the
ends of the track \cite{ZHAireS,ZHS_time}. To calculate travel times in ZHAireS we perform a
numerical integration to account for the variation of the index of refraction with altitude
assuming that the emission travels in straight lines to the observer (see below).
The curvature of the Earth's atmosphere is fully accounted for in AIRES. 

We have modified the ZHAireS code to deal with the reflection of air shower radio emission on a surface. 
For each rectilinear track element we first find the point on the reflection surface and the angle
of emission of radiation with respect to the track so that the emitted ray propagates first to the reflection point 
and then upwards towards the observer at a fixed position.
Approximating the reflection surface to a plane makes it trivial to obtain this
reflection point for a ray coming from any point in the atmosphere.
Once this is known, the time delay due to the refractive index is 
easily calculated integrating the travel time 
over the total path of the ray before and after the reflection.
We assume that the emission travels in a straight lines to the observer. 
We have explored the validity of this approximation using a simple one
dimensional model to produce pulses which resemble the fully simulated
ones. This model allows the calculation of travel times using numerical methods to
propagate signals along curved trajectories. The properties of the pulses obtained with straight
ray approximation are equivalent to those accounting for curvature up to
shower zenith angles $\theta\sim85^\circ$ and frequencies of $\sim 1$ GHz. 
This study is described in detail in the Appendix.

As mentioned before, we approximate the reflection surface to the $xy$-plane defined in Section~\ref{section2}, 
assumed to be perfectly flat.
The bulk of the emission has been shown to be concentrated in a cone that makes a small ``off-axis" angle $\psi$ 
to the shower direction as shown in Fig.~\ref{fig:psi}. This angle is very close to the Cherenkov angle 
($0.5^\circ-1^\circ$) at an altitude at which shower maximum occurs \cite{ZHAireS_ANITA}.
As a consequence the illuminated region on the reflective surface is 
relatively small, of order $0.5~{\rm km}$ $\times$ $1~{\rm km}$ ($1.5~{\rm km}$ $\times$ $10~{\rm km}$) 
for a $\theta=60^\circ$ ($80^\circ$) shower. 
As a result it is reasonable to ignore 
the differences in the orientation angle and altitude of the reflecting surface 
at the locations of the different reflection points across the illuminated area.
These differences are below $0.1^\circ$ and a few meters respectively even for showers 
of $\theta = 80^\circ$. 
There are other important aspects to the flat mirror approximation. 
When rays are reflected on a convex and rough surface 
they will diverge after reflection, and therefore the power received at a 
given surface element will be typically less than when a flat reflector is assumed.
These effects have been studied in~\cite{ref:SWORD} and~\cite{JupiterProbe}, 
and have little impact at moderate zenith angles, however they can become 
significant for high zenith angles ($\theta > 80^{\circ}$). 
Results of the simulations with the reflective surface assumed flat
can be corrected ``a posteriori" following the procedures outlined 
in ~\cite{ref:SWORD,JupiterProbe}. 
Such calculation is however very detector specific, and 
out of the scope of this article. 
Despite this, we expect the simulation method presented in this work
to be very suitable for that purpose. 

At each reflection point the Fresnel coefficients are applied to the
time-domain electric field to calculate the attenuation
of the components with polarization parallel, $r_\mathrm{\|}$, and 
perpendicular, $r_{\perp}$, to the reflection plane, defined by the normal to the reflecting 
surface and the direction of the radiation: 
\begin{equation}
r_{\perp} = 
\frac{n_1\cos\theta-n_2\sqrt{1-\left(\frac{n_1}{n_2} \sin\theta\right)^2}}
{n_1\cos\theta+n_2\sqrt{1-\left(\frac{n_1}{n_2} \sin\theta\right)^2}},
~~{\rm and}~~r_\mathrm{\|} =\frac{n_1\sqrt{1-\left(\frac{n_1}{n_2}
        \sin\theta\right)^2}-n_2\cos\theta}{n_1\sqrt{1-\left(\frac{n_1}{n_2}
        \sin\theta\right)^2}+n_2\cos\theta}.
\end{equation}
\begin{figure}
\begin{center}
\includegraphics[width=0.8\textwidth]{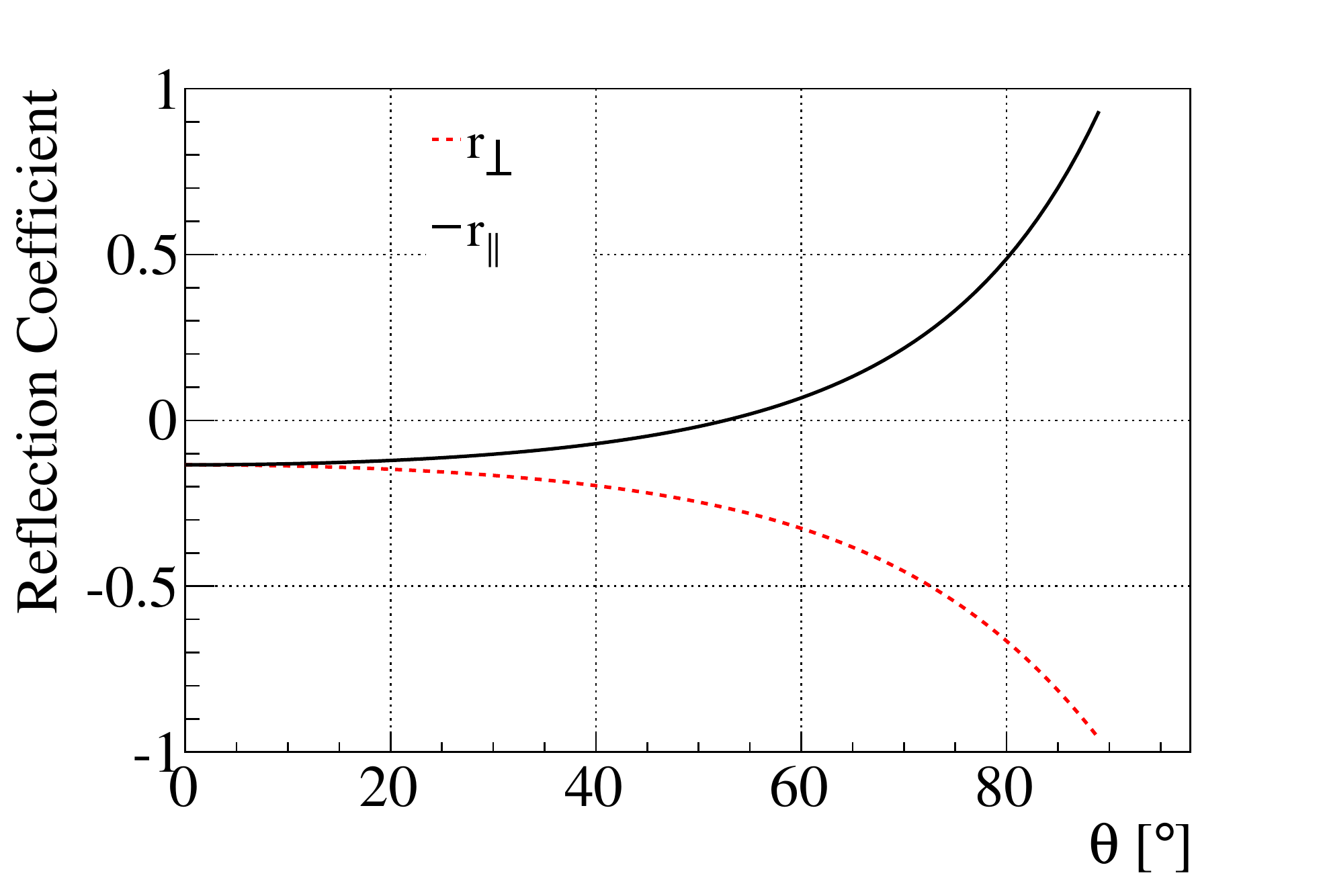}
\caption{The Fresnel coefficients for an air-ice interface 
with refractive indices 1.0003 and 1.31 respectively,
as a function of the zenith angle $\theta$ of the incident ray.
}
\label{fig:fresnel}
\end{center}
\end{figure}
The Fresnel coefficients for an air-ice interface are shown in Fig.~\ref{fig:fresnel} 
as a function of zenith angle $\theta$ of the incident ray. A large fraction of the components of the field 
is not reflected below $\theta\sim70^\circ-80^\circ$, 
and in fact at the Brewster angle at $\theta\sim 53^\circ$ 
the parallel component is not reflected at all. 
The coefficients change rapidly above $\theta\sim60^\circ$. 
Clearly they have a drastic impact on the 
the overall amplitude, the polarization and the zenith angle
dependence of the radio signal as will be shown in Section~\ref{sec:sims}.

\section{Simulations for a high altitude detector}
\label{sec:sims} 

\subsection{Simulation set and considerations}

In this section we apply the implementation of the reflection in ZHAireS to a
configuration that resembles a high altitude balloon experiment over Antarctica,
such as the ANITA flights, or the planned EVA mission \cite{EVA}.

We place antennas at a fixed altitude $h_d=36~{\rm km}$ above sea-level, 
and choose the reflecting surface to be at $z_g=2~{\rm km}$ above sea-level. We adopt 
a refractive index of $n =1.31$ \cite{Besson_ref_index}
consistent with ANITA measurements of the reflected image of the Sun~\cite{Stockham}. 
The geomagnetic field is chosen to have a typical value\footnote{See for example 
http://www.ngdc.noaa.gov/geomag/geomag.shtml} of 55\,$\mu$T and an
inclination of $-70^{\circ}$.
We generated proton showers with zenith angles 
$\theta = \{57^{\circ},64^{\circ},71^{\circ},78^{\circ},85^{\circ}\}$
and azimuth such that they always arrive from the geomagnetic west. 
For each zenith angle, we generate air showers with energies
$E = \{10^{17.8}, 10^{18.4},10^{19},10^{19.6}\}$~eV. 
We select simulations that have a  $X_{\text{max}}$ similar 
to the average $\langle X_{\text{max}}\rangle_{\text{Auger}}$ observed at the
Pierre Auger Observatory. To do so, we pre-simulate seven air showers per 
configuration with different random seeds and we 
select the air shower closest to $\langle X_{\text{max}}\rangle_{\text{Auger}}$.  
This results in an average deviation of $|X_{\text{max}} - \langle X_{\text{max}}\rangle_{\text{Auger}}|
\approx18$\,g\,cm$^{-2}$, which is within the root mean square of 
the energy-dependent $X_{\text{max}}$-distributions that have been 
observed. The shower simulation is run with AIRES using QGSJETII.03 hadronic
model interactions with a thinning level of $10^{-5}$~\cite{ref:Aires}. 

\subsection{Results}


%
\begin{figure}
\begin{center}
{
\includegraphics[width=0.47\textwidth]{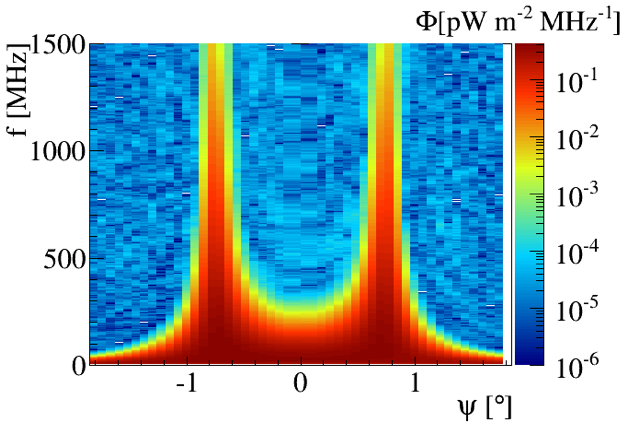}
\includegraphics[width=0.47\textwidth]{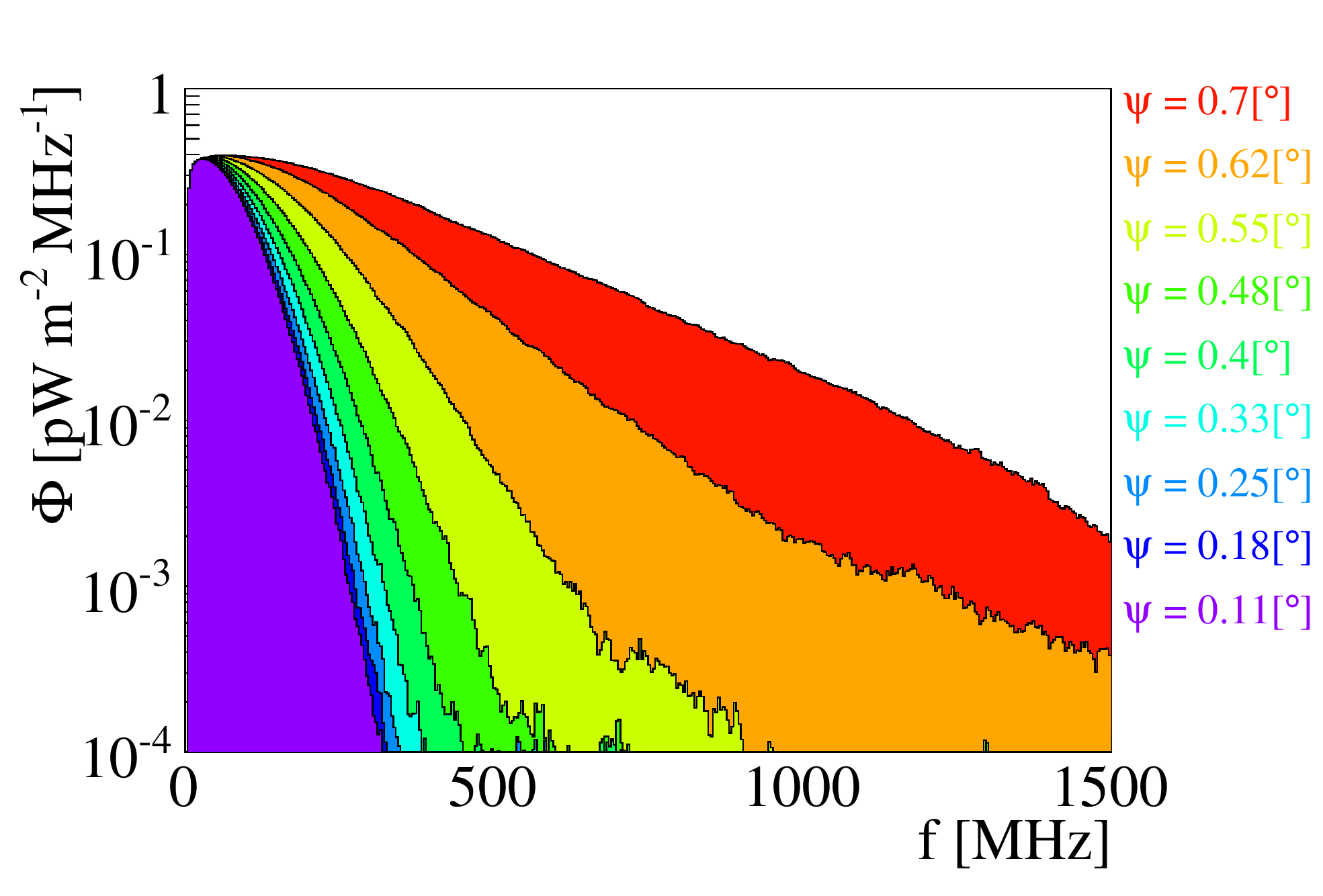}
}
{
\includegraphics[width=0.47\textwidth]{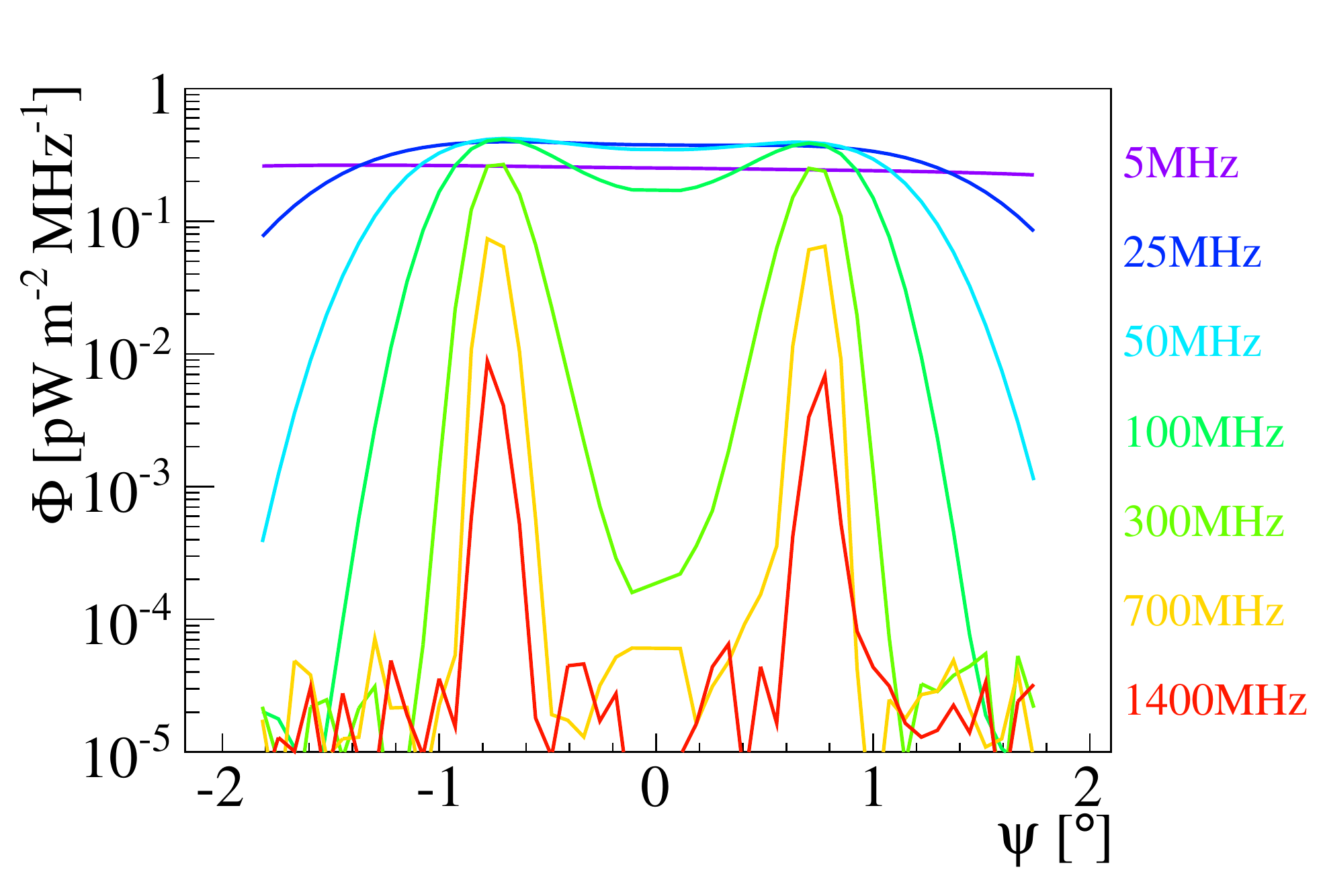}
\includegraphics[width=0.47\textwidth]{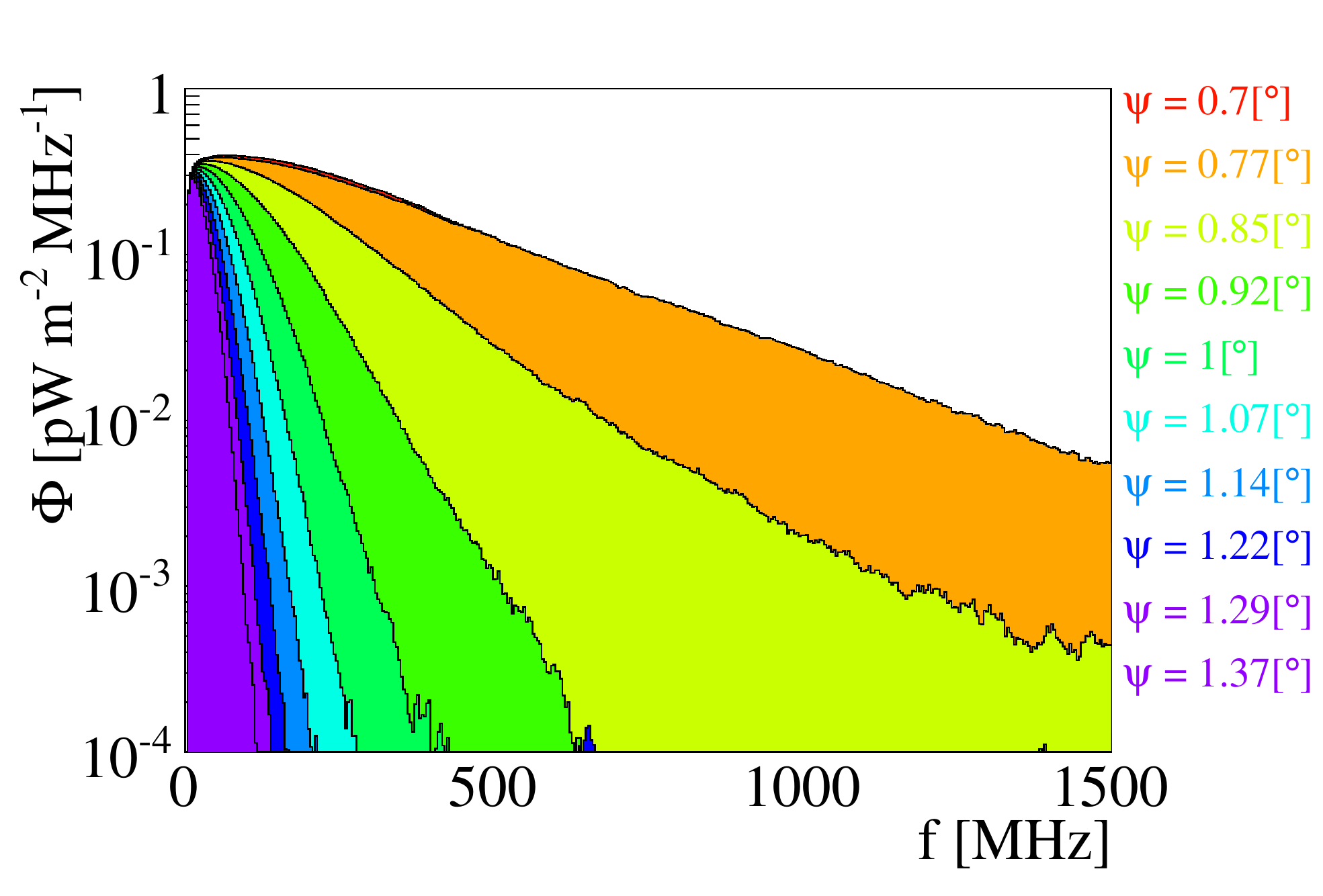}
}
\end{center}
\caption{
  Top left panel: distribution of the radio
  signal (flux density $\Phi$) as a function of off-axis angle $\psi$ (Fig.~\ref{fig:psi})
  and frequency $f$ for an air shower with
  $\theta = 71^{\circ}$ and $\log_{10}(E/\text{eV}) = 18.4$.
  In the bottom left panel we show the
  distribution of the radio signal as a function of off-axis angle at various frequencies. 
  In the right panels we show the radio signal
  distribution as a function of frequency, in the top right panel 
  for off-axis angles equal or smaller than the Cherenkov-angle at the $X_{\rm max}$ of the shower
  i.e. $\psi \leq 0.7^{\circ}$, while in the bottom right panel for $\psi \geq 0.7^{\circ}$.
  }
\label{fig:sigDist}
\end{figure}
To illustrate some of the typical features of the radio signal, we display 
in Fig.~\ref{fig:sigDist} the flux density $\Phi$ as a function 
of frequency and off-axis angle $\psi$ for an air shower with zenith angle
$\theta=71^{\circ}$ and a energy $E=10^{17.8}$~eV. The flux density in this case 
is defined as the power spectrum at a fixed frequency $f$ averaged over a period of 10\,ns,
and is given in units of ${\rm pW~m^{-2}~MHz^{-1}}$ throughout this paper. 
In the top left panel of Fig.~\ref{fig:sigDist} we show the two-dimensional distribution as a function of $\psi$ and $f$ 
which displays coherent properties and is clearly beamed around the Cherenkov angle at $\sim 0.77 ^\circ$. 
This can be better appreciated in the bottom left panel where we show the off-axis distributions 
for different frequency components of the pulse. As the frequency increases 
the radiation adds coherently only within a smaller angle off the Cherenkov cone.
In the right panels in Fig.~\ref{fig:sigDist} we show the spectral
shape of the flux density for a variety of observation off-axis angles. 
At very low frequencies ($f<10~{\rm MHz}$) the flux density increases until 
it reaches a maximum in the range ($f\sim 10 - 150~{\rm MHz}$) and then decreases with an
exponential fall-off to first order. 
A very important feature is illustrated in the right panels, the steepness of the
fall-off has a clear dependence on the off-axis angle $\psi$ of the detector. 
This dependence is key to the energy determination of UHECRs 
with ANITA as pointed out in~\cite{BelovARENA}. 

\subsubsection{Implications of the reflection}

Other efforts to simulate reflected radio signals from air showers have relied on pulses
simulated at ground which were extrapolated using the attenuation
of the signal with increased distance ($|\vec{E}| \propto 1/r$)
after accounting for the loss of signal induced by the Fresnel coefficients \cite{BelovARENA}. 
In a homogeneous medium this ``specular approach" can be expected 
to be a good approximation provided the pulse can be considered to be in the
Fraunhofer limit. As a result it can be expected to work better for highly
inclined showers, since the distance between the observer and air shower
increases as the zenith angle rises.

In this work all track contributions to the pulse are reflected at the
interface to account for attenuation with distance, for the Fresnel-reflection
coefficents attenuating the parallel and perpendicular components of the field, 
and for the fact that reflection also alters the relative time delays of emission 
from different regions of the shower affecting the coherence properties of the
pulses. Therefore, this method can also be applied 
when the reflector is not in the Fraunhofer limit. 

\begin{figure}
\begin{center}
\includegraphics[width=0.47\textwidth]{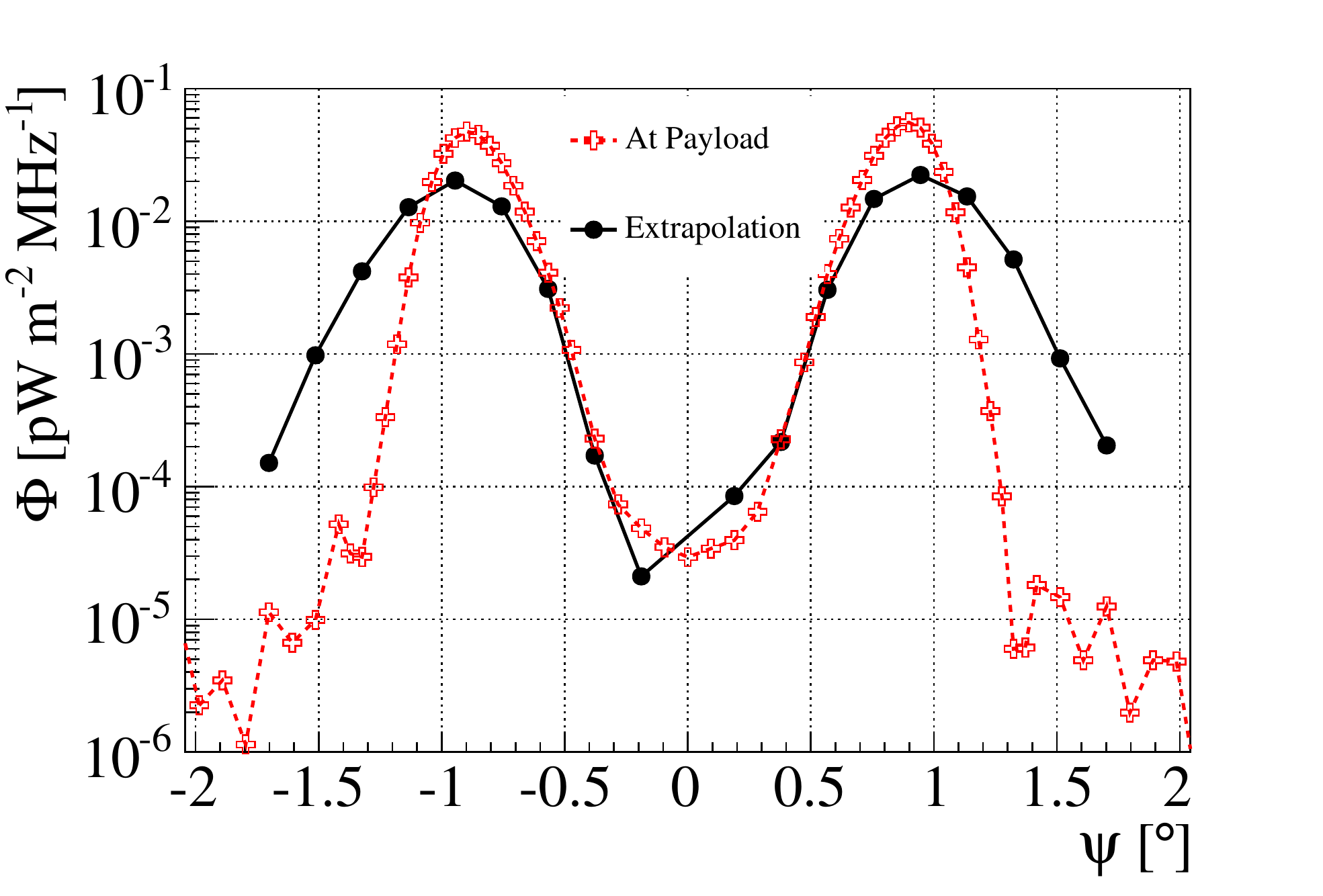}
\includegraphics[width=0.47\textwidth]{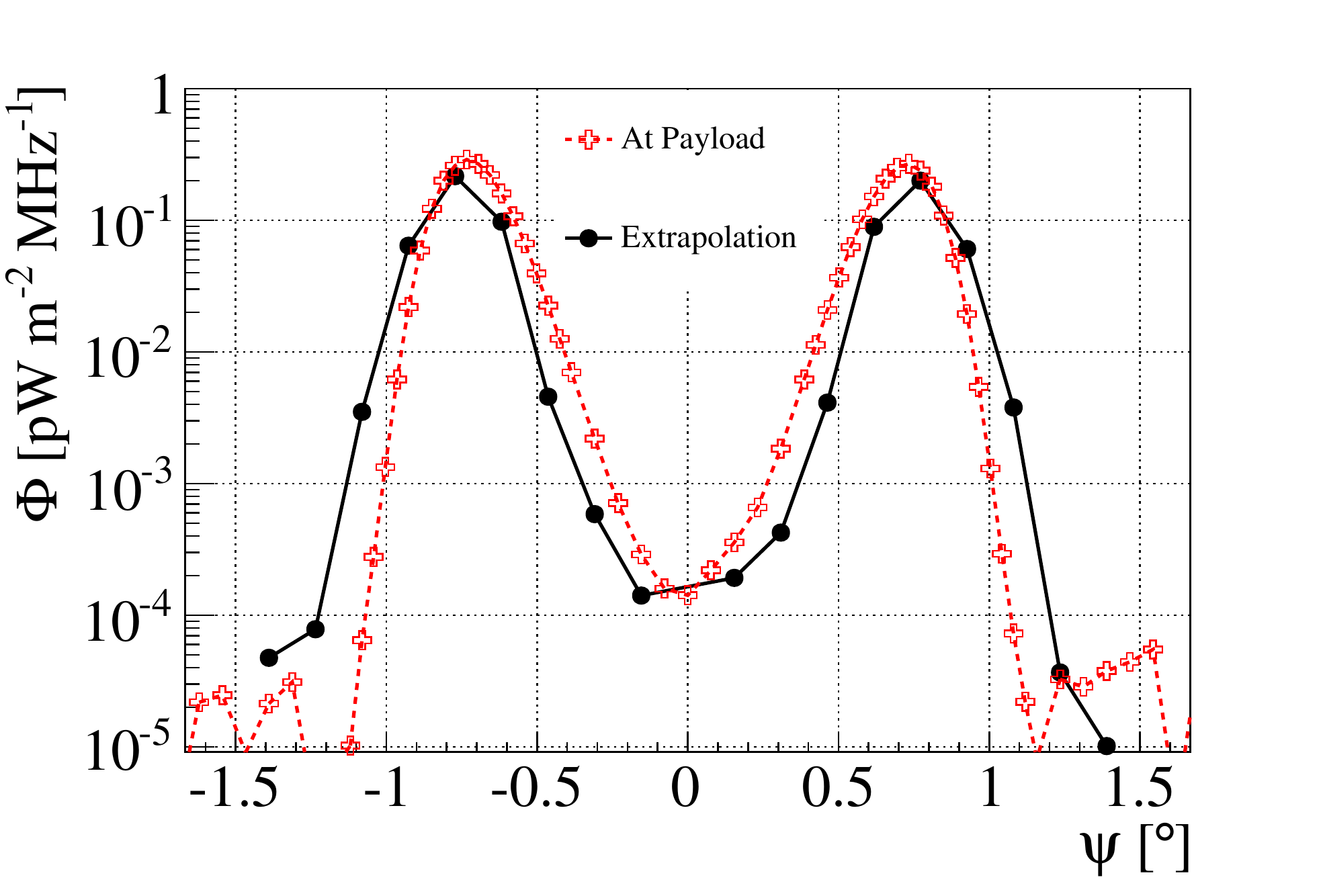}
\caption{Flux density $\Phi$ at a frequency $f=300~{\rm MHz}$ as a function of off-axis angle $\psi$ as obtained
 extrapolating the ZHAireS simulated signal at ground to the detector (black dots and solid line), 
and simulating the reflection as explained in the text (red crosses and dashed line). The flux density
is shown for showers of $E=10^{18.4}$ eV and two zenith angles $\theta=57^{\circ}$ (left panel) 
and $\theta=71^{\circ}$ (right panel). 
The Fresnel-reflection coefficients are accounted for in all cases.
}
\label{fig:GroundVsAnita}
\end{center}
\end{figure}

We compare the specular approximation to the results of the full ZHAireS simulation
including reflection to illustrate the difference between the two methods.
For this comparison we evaluate the flux density at $f=300~{\rm MHz}$ at a few
ground locations scaling it to account for the distance from 
ground to the location of the high altitude balloon. 
In Fig.~\ref{fig:GroundVsAnita}
we display the flux density as a function of $\psi$ for two
different zenith angles. 
We note that for $\theta=57^{\circ}$ (Fig.~\ref{fig:GroundVsAnita} left) the distribution 
in $\psi$ is significantly wider what can lead to orders of magnitude of
over-estimation of the flux density for the larger off-axis angles. 
For $\theta=71^{\circ}$ (Fig.~\ref{fig:GroundVsAnita} right) we still see relevant deviations between the two
methods, but they are significantly reduced compared to the lower zenith angle case.
Significant deviations are also found at other frequencies. Moreover, the 
shapes of the frequency spectra obtained with the two methods also differ appreciably.

\begin{figure}
\begin{center}
\includegraphics[width=0.47\textwidth]{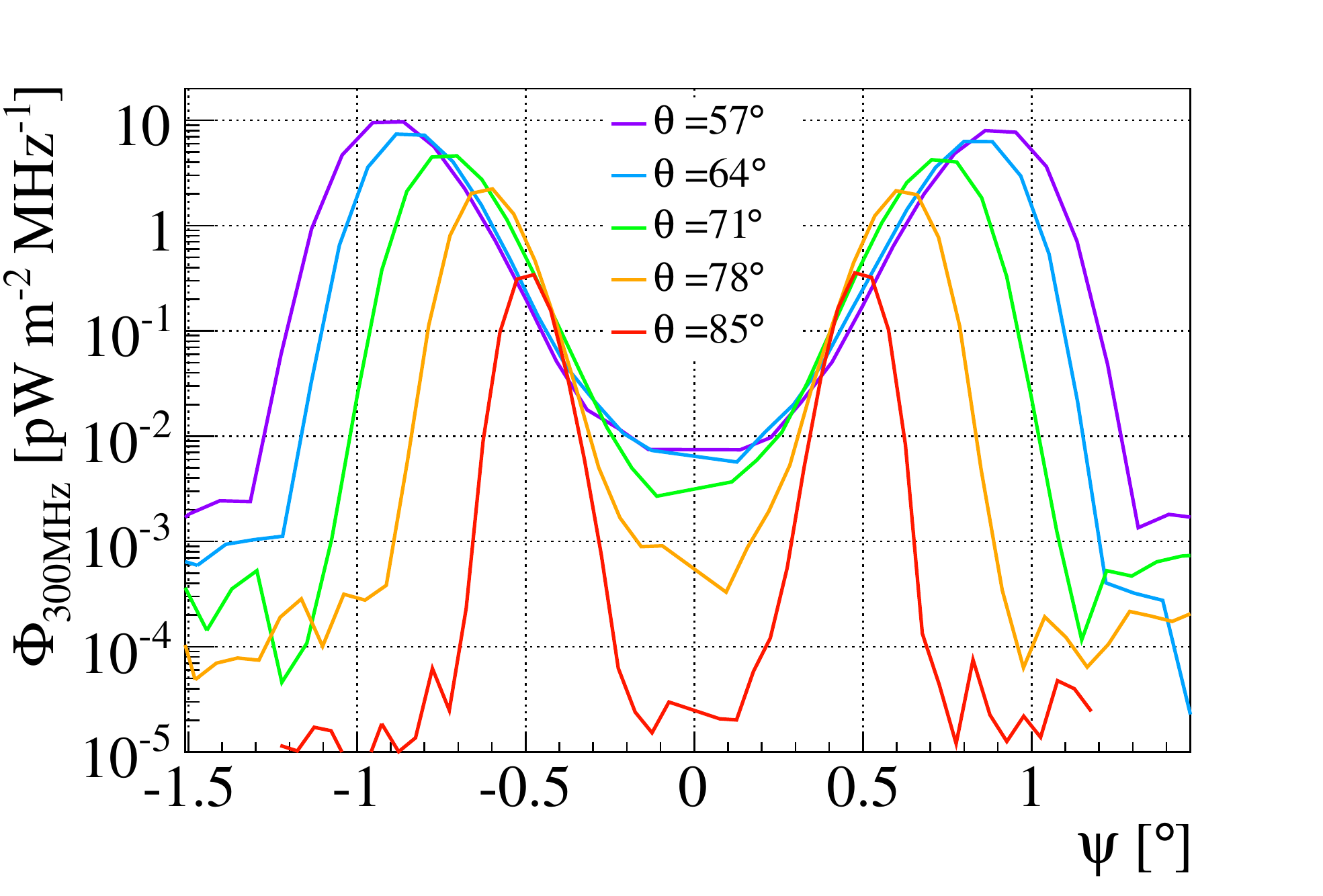}
\includegraphics[width=0.47\textwidth]{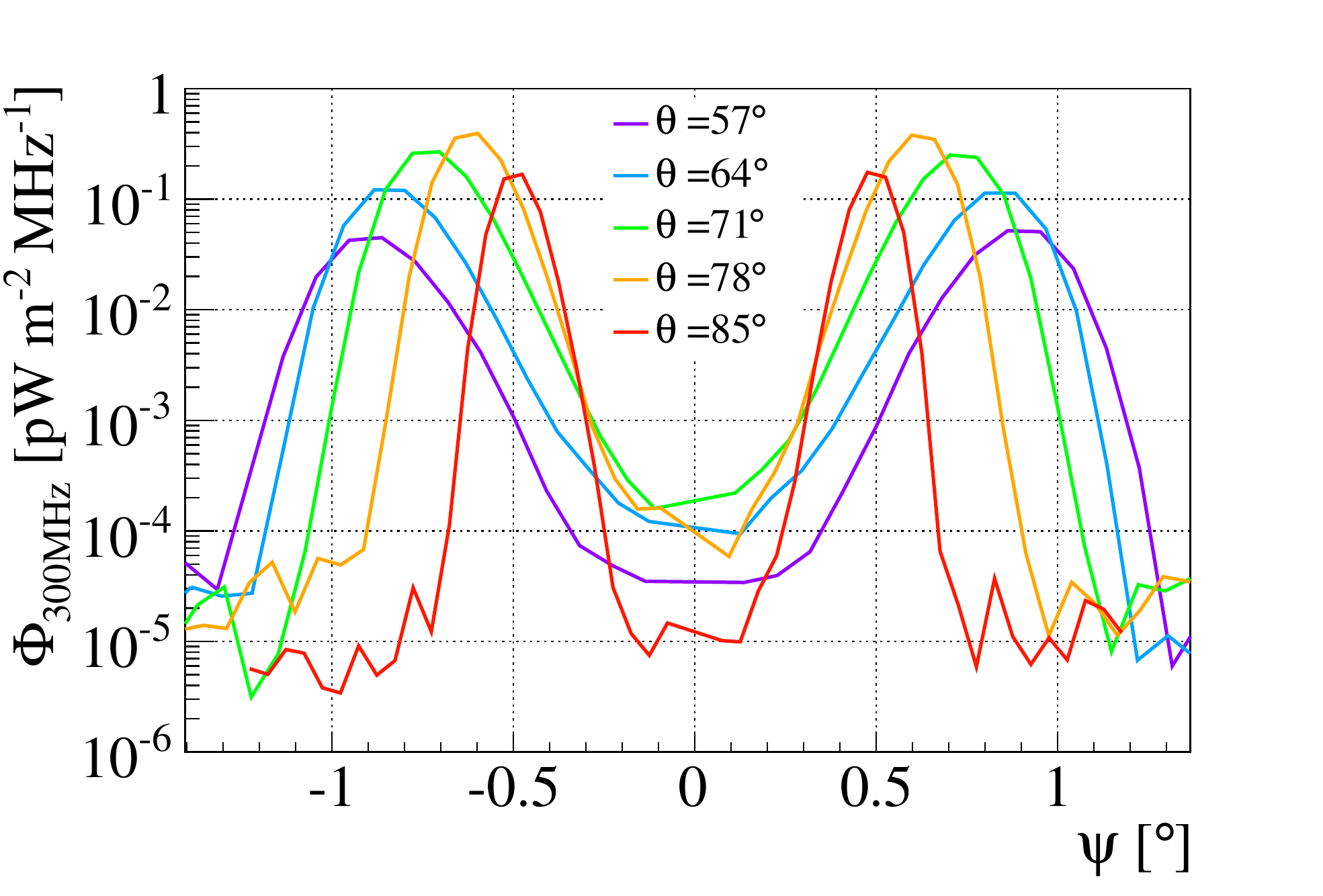}
\caption{Comparison of the flux density $\Phi$ at a frequency $f=300~{\rm MHz}$ as a function of
the off-axis angle $\psi$ before (left) and after applying Fresnel
  reflection coefficients (right). Different sets of curves correspond to
  different shower zenith angles $\theta$ as labeled. The simulated showers have an energy $E=10^{18.4}$\,eV. }
\label{fig:300Off-axis}
\end{center}
\end{figure}

It is interesting to explore how the radiation changes with zenith
angle for a primary particle with fixed energy. As the zenith angle increases 
the flux density $\Phi$ decreases.
This can be seen in the left panel of Fig.~\ref{fig:300Off-axis}, displaying 
$\Phi$ at $f=300~{\rm MHz}$ as a function of $\psi$ for an air shower 
induced by a primary particle with energy $E = 10^{18.4}$~eV.
The dominant effect in the decrease is the increasing overall
distance to the detector with $\theta$ (see Fig.~\ref{fig:dXmax}). 
Other effects however compensate the decrease in $\Phi$. 
The angle $\alpha$ of the shower axis to the Earth's magnetic field at the South Pole 
increases in the range of $\theta$ shown in Fig.~\ref{fig:300Off-axis},
and the geomagnetic contribution is known to scale with $\sin\alpha$.  
Also showers of increasing $\theta$ develop in a less dense
atmosphere where the geomagnetic contribution to the electric field is expected
to be increasingly larger \cite{Scholten_MGMR}.
The net result, including
other more subtle effects \cite{Washington_ARENA12}, 
is a decrease of $\Phi$ with $\theta$.
 
To illustrate the importance of accounting for the Fresnel reflection coefficients 
they were artificially set to 1
in the simulations shown in the left panel of Fig.~\ref{fig:300Off-axis}, 
while in the right panel they are taken into account. 
Comparing both panels, 
the peak value of the flux density is largest at relatively 
high zenith angles ($\theta \sim 80^\circ$)
when the Fresnel coefficients are accounted for,
contrary to what is seen in the left panel where the peak value
of $\Phi$ is achieved at the smallest zenith angles. This suggests that detection can be 
expected to be most favorable for $\theta$ around $80^\circ$. A thorough calculation of the
acceptance integrating over area and solid angle~\cite{ref:Motloch} should
also account for the reduction of the Cherenkov angle as the
zenith angle rises (see Fig.~\ref{fig:dXmax}), and for the directionality 
of the detection system. Such calculation is out of the scope of this article. 

\subsubsection{Energy dependence}

From the set of simulations we have examined the energy dependence of
the radio signal. As before, we use the flux density at a reference
frequency $f=300~{\rm MHz}$ for a shower of $\theta=71^{\circ}$. 

\begin{figure}
\begin{center}
\includegraphics[width=0.8\textwidth]{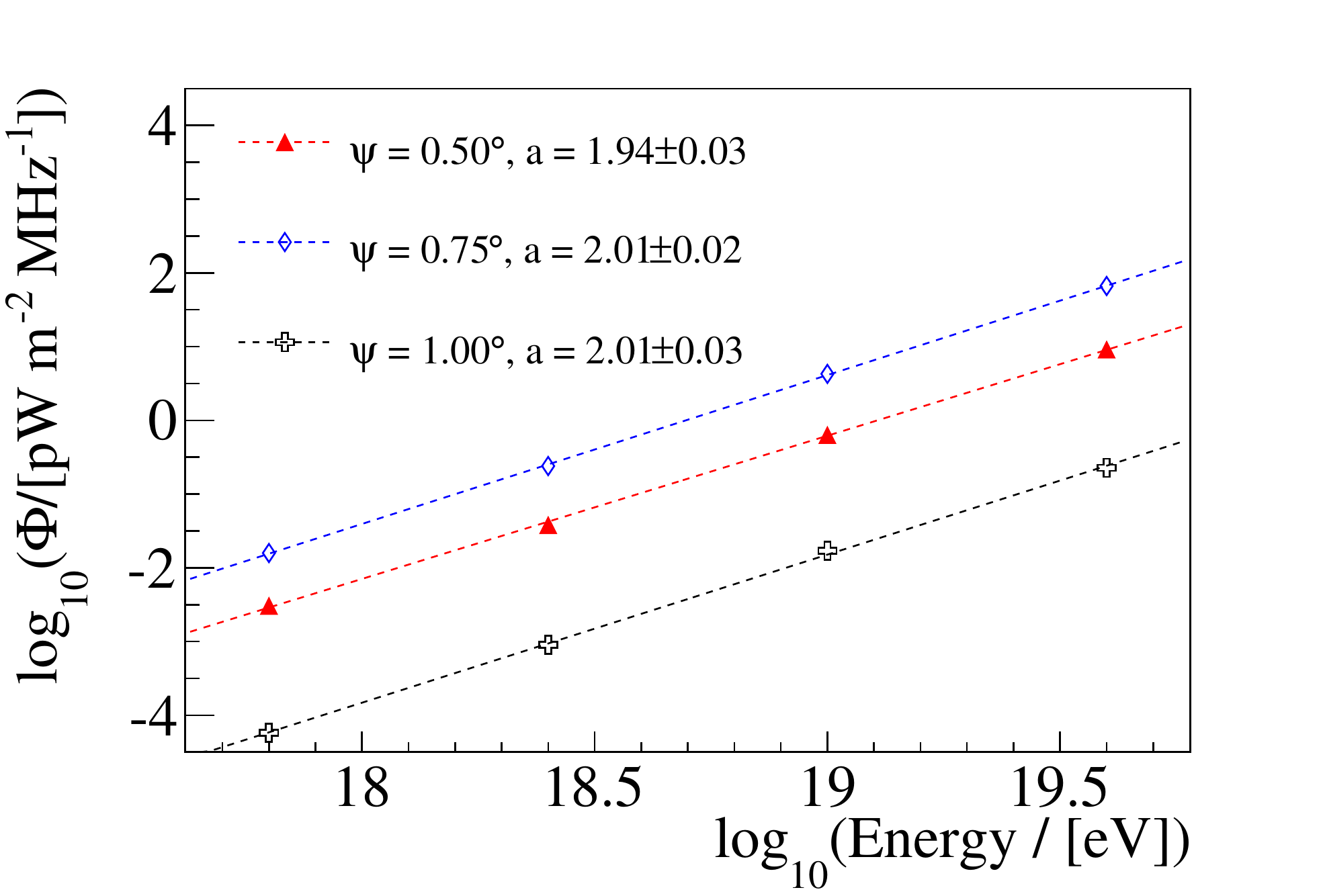}
\caption{The flux density $\Phi$ at frequency $f=$300\,MHz as a
 function of the energy of the primary particle for three off-axis
 angles in a shower of $\theta=71^\circ$. 
The results of fitting a straight line $\log_{10} \Phi =  a
 \log_{10}(E/\text{eV}) + b$ are shown.}
\label{fig:Energy}
\end{center}
\end{figure}
In Fig.~\ref{fig:Energy} we select three
off-axis angles and plot the flux density as a function of the 
primary particle energy. We fit a simple linear function to the
dependence of $\log_{10}\Phi$ on $\log_{10} (E)$ and find a slope
that is consistent with 2. This confirms that the received flux
density scales quadratically with the primary energy and 
the amplitude of the electric field scales linearly with it. 
This is not surprising since for a coherent signal it is expected 
that the amplitude of the electric field scales with 
the number of electrons in the shower which is proportional to the 
energy of the primary particle. 
We verified that the quadratic relation
between flux density and energy of primary cosmic ray particle holds 
for all the zenith angles 
and all considered frequencies in our simulation set. 
This has important consequences since measuring 
the flux density at a given off-axis angle provides a 
measurement of the energy of the air shower. 
In practice the off-axis angle can be
related to the exponential fall-off of the flux density as
can be seen in Fig.~\ref{fig:sigDist}.
This relation is key to the energy determination 
of UHECRs with ANITA \cite{BelovARENA}. 
This means that it is in principle possible to deduce
the primary energy from a single location as long as the 
exponential fall-off of the spectrum can be determined.

\section{Conclusions}

In this paper we implemented the treatment of surface reflection for
radio signals from air showers, upgrading existing ZHAireS simulations\footnote{Code available upon request.}. 

As a case study we simulated radiation from a set of air showers 
at a high altitude position over Antarctica inspired by 
the ANITA experiment which has provided the only measurements 
of reflected radio pulses from air showers up to now. 
We described and explained the flux density distributions 
as a function of frequency, energy and 
zenith angle of the primary cosmic ray particle. 
We have stressed the importance of accounting for the Fresnel-reflection 
coefficients attenuating the components of the field, as well as accounting   
for the fact that reflection also alters the relative time delays of emission 
from different regions of the shower affecting the coherence properties of the
pulses.

A clear quadratic correlation between the radio observable (flux density) and
the shower energy has been observed in ZHAireS simulations with reflection. 
The intercept of the correlation depends on the off-axis angle which can 
be related to the exponential fall-off of the spectral distribution of the flux density.
This provides the basis for the determination of shower energy from a single location \cite{BelovARENA}. 

Several approximations were made in the implementation of the reflection in the
ZHAireS code. The reflective surface is assumed to be a plane and we argued 
that this is a good approximation as long as the shower zenith angle is $\theta\lesssim80^\circ$.
Also the emitted radiation is assumed to travel along straight lines to the ground
and then to the detector, an approximation that has been extensively verified in 
Appendix A. 

Finally, we foresee many applications of this code to the design and
feasibility studies of future large exposure cosmic ray detectors
based on the radio technique applied to air showers~\cite{ref:SWORD,ref:Motloch}.

\section*{Acknowledgements}
We thank Ministerio de Econom\'\i a (FPA2012-39489), 
Consolider-Ingenio 2010 CPAN Programme (CSD2007-00042),  
Xunta de Galicia (GRC2013-024), Feder Funds  
and Marie Curie-IRSES/ EPLANET (European Particle physics Latin
American NETwork), $7^{\rm th}$ Framework Program (PIRSES-
2009-GA-246806). H.S. is supported by Office of Science,
U.S. Department of Energy and N.A.S.A. 
We also thank CESGA (Centro de Supercomputaci\'on de Galicia) for computing resources.


\appendix

\renewcommand*{\thesection}{\Alph{section}}
\section{Appendix}
\label{sec:RayVsStraight}

\subsection{Straight vs curved ray propagation}

The variation with altitude of the index of refraction of the atmosphere
is known to induce curvature in the path of the radio waves when propagating.
It has been shown that the effects are negligible for most shower geometries
and observers on ground \cite{werner08}, for which the propagation along straight paths is a good
approximation.    
In the case of showers at large zenith angles and especially when accounting
for reflection, the involved distances from emission
to the detector become large (see Figs.~\ref{fig:geometry} and \ref{fig:dXmax}), 
and the curvature of the rays can be expected to increase. 

To evaluate if the approximation of straight light ray propagation still holds
in the typical geometries involved in reflection, we have 
developed a simple ray tracing code. We divide the atmosphere in many 
layers with constant distance between them. 
The layers are taken sufficiently
narrow so that the ray can be approximated as traveling in a straight line along
a constant refractive index $n$ in each layer
given by the exponential model in Section \ref{section2}. 
The ray is refracted at each interface,
taking into account the different refractive index at each layer.
The travel time 
of the ray is calculated as the sum of the times it takes to cross each layer.  
When accounting for reflection on the ground, the ray is  
also propagated upwards through a decreasing refractive index profile 
until it reaches the detector. 
The arrival time assuming straight line propagation
to ground and then to the same detector position is also calculated.
The reflection surface is assumed to be at sea level for
these calculations. Since the gradient of an exponential atmosphere is largest at
sea level, it can be expected that curvature effects for reflection from surfaces at higher
altitudes will have less impact than estimated here.

\begin{figure}[htbp]
\begin{center}
{
\includegraphics[width=0.7\textwidth]{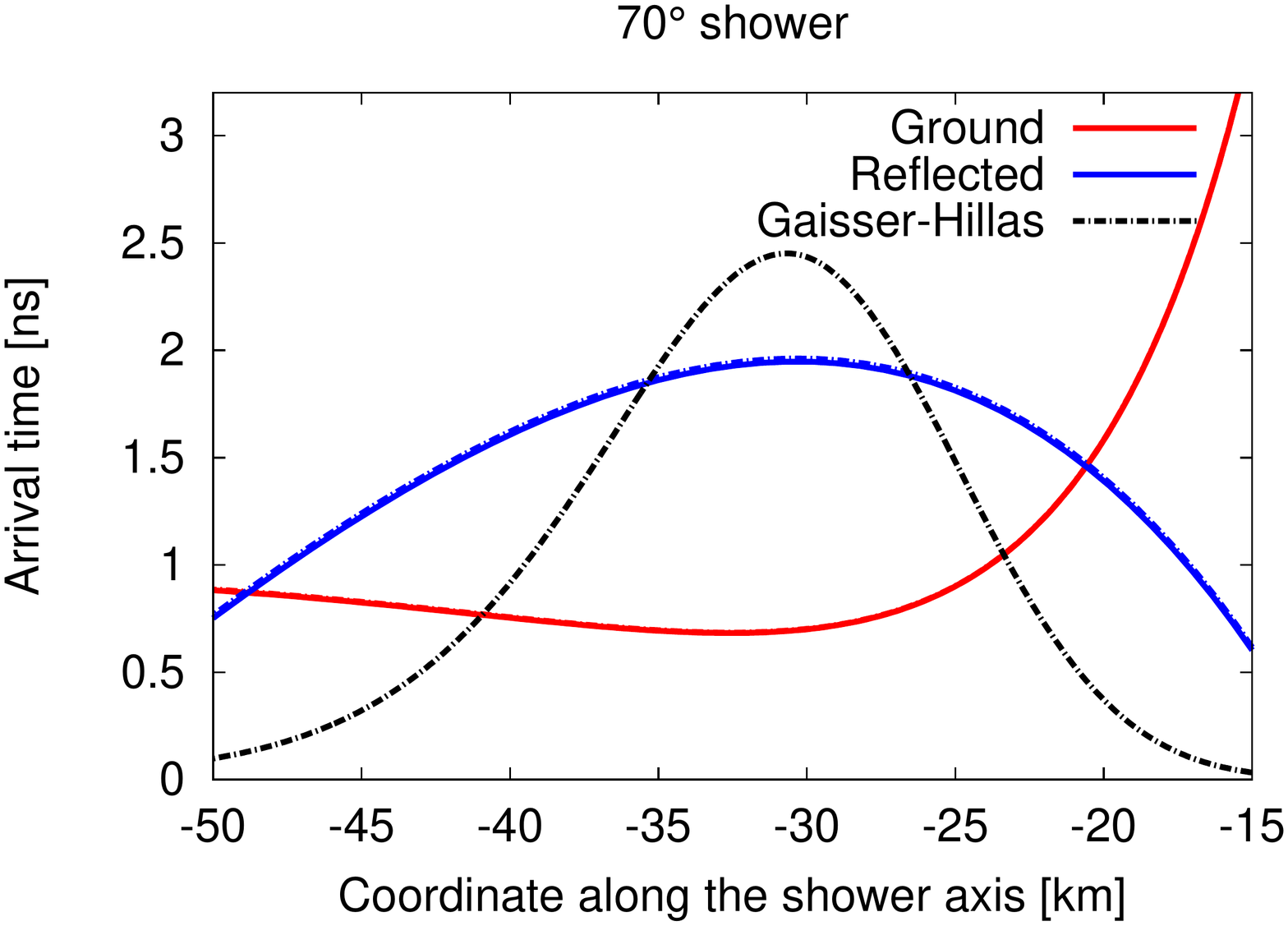} \\
\includegraphics[width=0.7\textwidth]{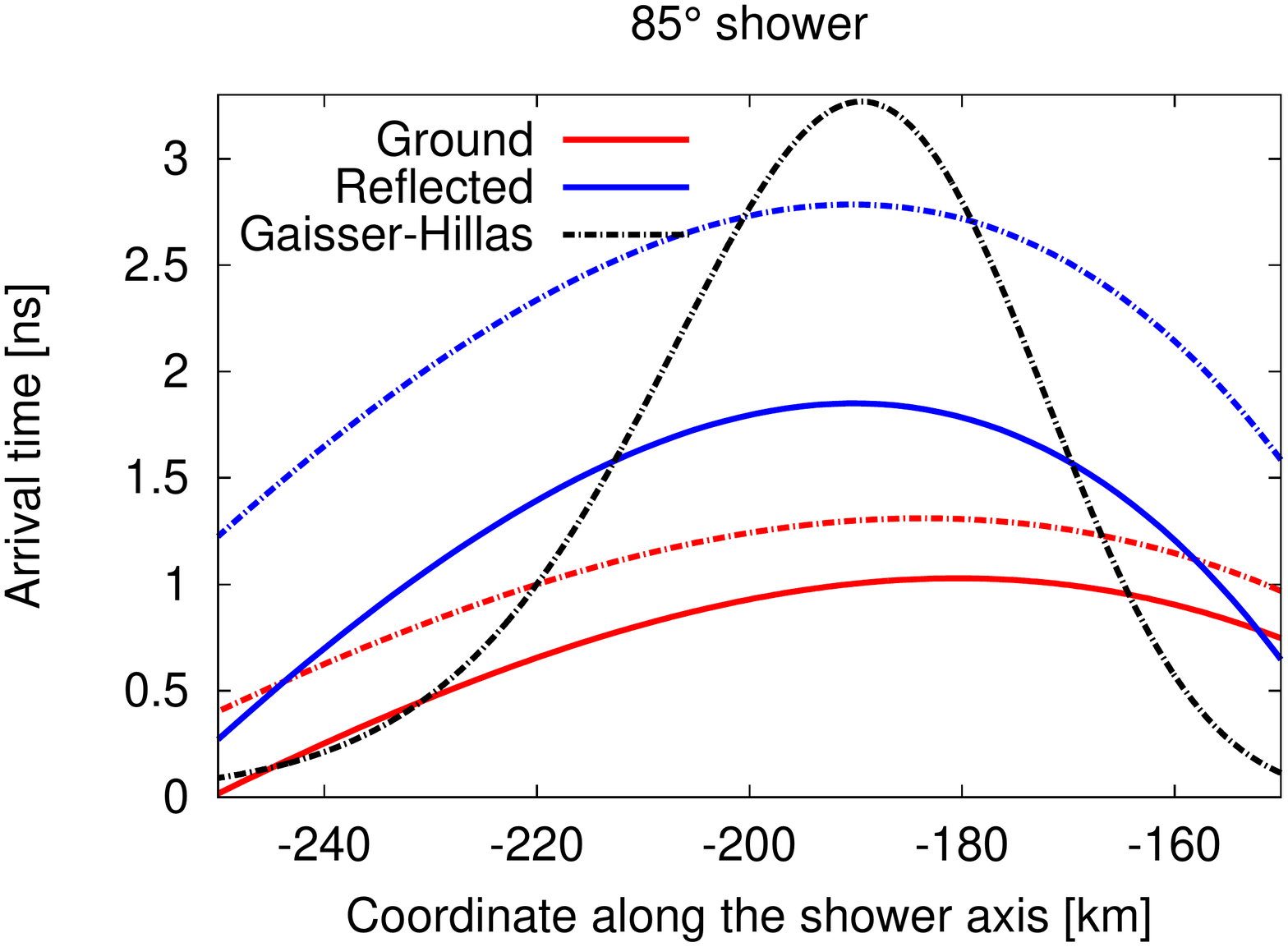}
}
\caption{
Top: Relative arrival time of rays emitted along the axis of a $\theta=70^\circ$ shower
at two particular antenna locations: one located on the ground (red lines) 
and another after reflection towards a high altitude balloon (blue lines)
both at an off-axis angle close to the Cherenkov angle $\psi\sim0.77^\circ$ (see Fig.~\ref{fig:psi}). 
The straight ray approximation (dashed lines) and the curved ray propagation (solid lines) are shown (see text for details). 
A $10^{19}$ eV, $\theta=70^\circ$ Gaisser-Hillas shower profile is superimposed (black line). 
Bottom: the same as in the top panel, but now for a $\theta=85^\circ$ shower viewed at an off-axis angle
$\psi\sim0.40^\circ$.}
\label{fig:times}
\end{center}
\end{figure}

In Fig.~\ref{fig:times} (top panel) 
we show the relative arrival times of radio signals emitted from different positions along the 
shower axis of a $\theta=70^\circ$ shower. They have been calculated with the straight 
and curved ray approximations  
for two particular observer positions such that the radiation arriving
from shower maximum makes an off-axis angle $\psi \sim 0.77^\circ$.
One observer is located on the ground and receives the
rays directly, the other observer is placed at an altitude $h_d \simeq 33~{\rm km}$ and 
receives the rays after reflection on the ground (see Fig.~\ref{fig:geometry}). 
The difference in the arrival times between curved and straight ray
propagation is hardly noticeable in the scale of Fig.~\ref{fig:times}. 
It is in fact below $\sim50$ ps,
which corresponds to a frequency of $\sim5$ GHz when using a quarter 
wavelength criterion for coherence. 
As a result we expect the straight ray approximation to be valid 
below this frequency.

Similarly in Fig.~\ref{fig:times} (bottom panel) we show the arrival times of
the pulses for a $\theta=85^\circ$ shower and observation at an off-axis angle 
$\psi=0.4^\circ$. Although in this case the difference between curved and
straight ray propagation is sizable,  
it is an almost constant offset along shower development 
of $\sim 0.3$ ns for the observer on the ground and 
$\sim 0.9$ ns for the observer at high altitude $h_d=50~{\rm km}$.
This global offsets induce unobservable phase shifts in the field at the detector. 
When accounting for these offsets,
the relative differences between the straight and curved propagations 
are $\sim 200$~ps for the observer at ground level
corresponding to a frequency of $\sim 1.25$~GHz using a quarter wavelength criterion, 
while for the observer at $h_d=50~{\rm km}$ altitude the differences are below $\sim20$ ps 
($\sim 12$ GHz frequency). 

In the top and bottom panels of Fig.~\ref{fig:times} the relative arrival times
at ground and at the high altitude observer 
are approximately flat within a large region 
($\sim 10~{\rm km}$) around shower maximum. As a result the emission from this region 
is coherent up to GHz frequencies.
An interesting feature in the top panel of Fig.~\ref{fig:times} is the inversion
of the arrival times at the high altitude observer position. The radio signal emitted 
in the Cherenkov angle (in this particular case from the region around shower maximum) 
arrives last at the high altitude location, contrary to what could be
expected in a homogeneous medium. 
The time inversion is also seen for the observer on
the ground for the $85^\circ$ shower in Fig.~\ref{fig:times} (bottom). 
The effect does not seem to have relevant implications for detection.   

\subsection{A simple model to describe the pulses}

The comparison of the travel times for the straight and curved ray
calculations already sheds light onto the validity of the straight
ray approximation. However a comparison between the pulses obtained 
with the straight and curved approaches in a simplified calculation 
will give us an estimate of the quantitative uncertainty 
in the properties  of the signal at the detector when using the
straight path approximation. 

For that purpose, we model a shower with zenith angle $\theta$ as a one-dimensional charge distribution 
varying with time $N(t)$ as the shower propagates along a given direction at the speed 
of light $c$~\cite{1Dmodel}. 
In the Fraunhofer approximation \cite{DanielZHS}, the electric field induced by this line of 
charge is proportional to the charge $N(t)$, to $r^{-1}(t)$, with $r(t)$ the distance between the emission point
and the observer position, and carries a phase factor to account for the time delay between
the arrival of the signal emitted from different positions along the line: 
\begin{equation}
E \sim \int \mathrm{d}t \ N(t) \frac{e^{i\omega t_a(t)}}{r(t)} 
\label{eq:minimalistic_field}
\end{equation}
Here $\omega$ is the angular frequency and $t_a$ is the arrival time at the
observer of a signal emitted at time $t$, where 
\begin{equation}
t_a(t) = t + \frac{n}{c}~r(t)
\end{equation}
This approximation was developed for a homogeneous, isotropic and non-conductive medium 
but it can be extended to the atmosphere accounting for the time delays from 
propagation of the rays in the altitude-dependent index of refraction~\cite{ZHAireS}. 

\begin{figure}[htbp]
\centering
{
\includegraphics[width=0.7\textwidth]{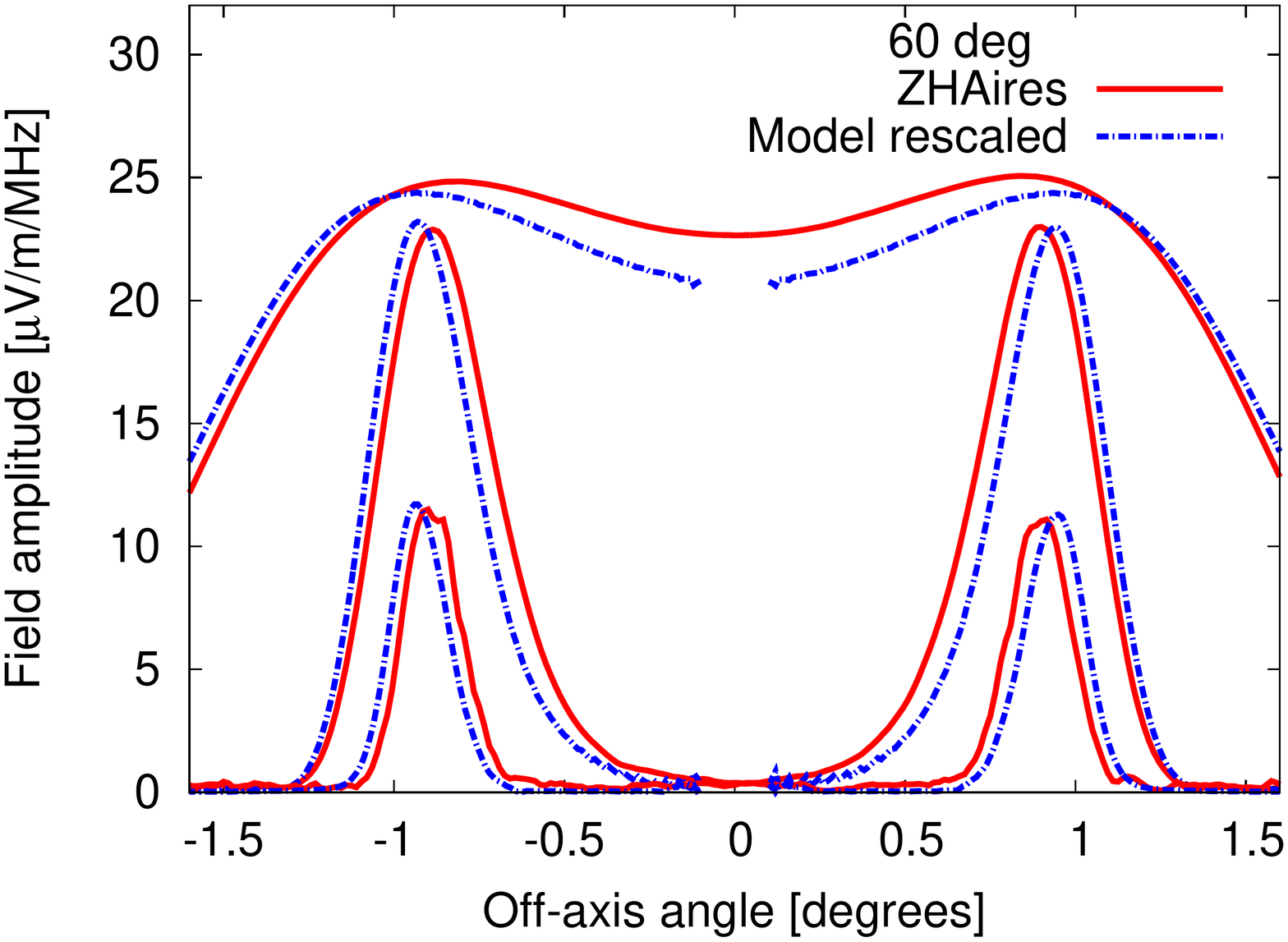}\\
\includegraphics[width=0.7\textwidth]{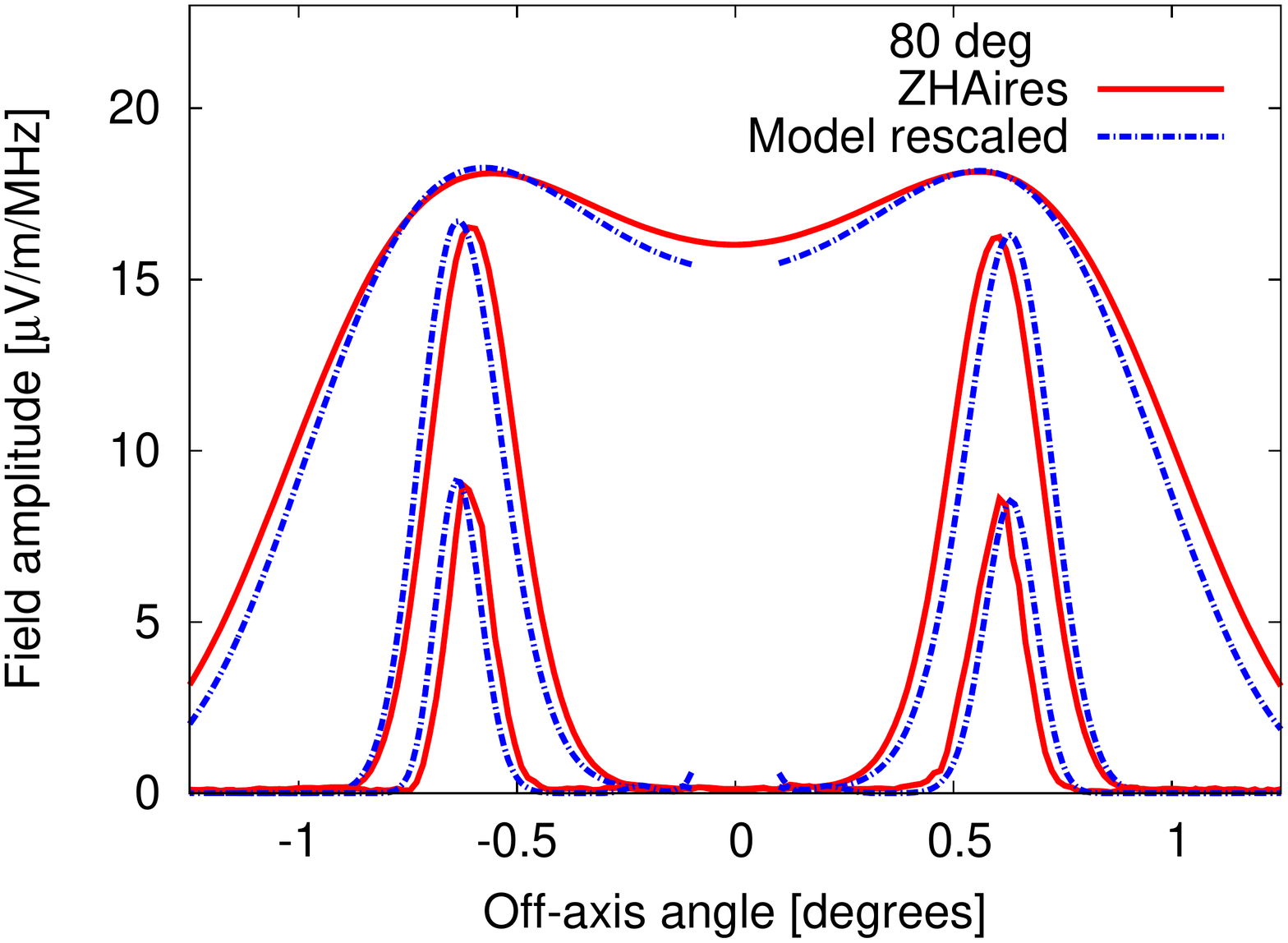}
}
\caption{Top: amplitude of the Fourier transformed electric field as a function of the off-axis angle
for different frequencies.
Results from full ZHAireS simulations and the model in Eq.~(\ref{eq:minimalistic_field}) are shown. 
The shower has $E=10^{19}$ eV, $\theta=60^\circ$ and a height of $X_{\rm max}$ above ground  $h_{\rm Xmax}\sim 4.4~{\rm km}$.
The observers are placed at a constant distance from $X_{\rm max}$ 
of $\sim 83.3~{\rm km}$. From top to bottom the 
observation frequencies are $50$, $300$ and $900~{\rm MHz}$. 
Bottom: Same as in the top panel, but for a $\theta=80^\circ$ shower with  
$h_{\rm Xmax}\sim 13.5~{\rm km}$ and with the observers placed at a constant distance 
of $\sim 276.9~{\rm km}$. 
}
\label{fig:sim_60_constdist}
\end{figure}

In Fig.~\ref{fig:sim_60_constdist} the pulses obtained in the full ZHAireS simulation
modified for reflection 
are compared to those obtained with Eq.~(\ref{eq:minimalistic_field}) using as input a 
one-dimensional Gaisser-Hillas distribution, $N(t)$, with the same 
depth of shower maximum as the simulated shower. 
The amplitude of the reflected field is shown for observers  
at a high altitude located at different off-axis angles $\psi$ (see Fig.~\ref{fig:psi}) 
and constant overall path distance. 
The results have been normalized to the peak of the
electric field as predicted with Eq.~(\ref{eq:minimalistic_field}). 
The shape of the angular distribution is well described by the simple model in a wide
frequency range. It should be noted here that, being one-dimensional, the approach can not
fully reproduce the frequency spectrum as obtained in the full simulations near
the Cherenkov angle, where the lateral spread is of utmost importance
\cite{ZHAireS}. We are however confident that the approximation is
sufficient to test the validity of the straight ray approximation. 

\subsection{Validity of the straight ray approximation}

We used the simplified one-dimensional model described above
to test the straight ray approximation. 
The arrival times $t_a$ at the detector to be used in Eq.~(\ref{eq:minimalistic_field}) 
are calculated integrating the travel time along both straight
and curved paths using the ray tracing algorithm explained above.
The electric field is obtained numerically, discretizing the integral in 
Eq.~\eqref{eq:minimalistic_field} in such a way that the time intervals
correspond to regions of the shower with approximately constant 
$N(t)$, $r(t)$ and $t_a(t)$. 

\begin{figure}[htbp]
\centering
{
\includegraphics[width=0.7\textwidth]{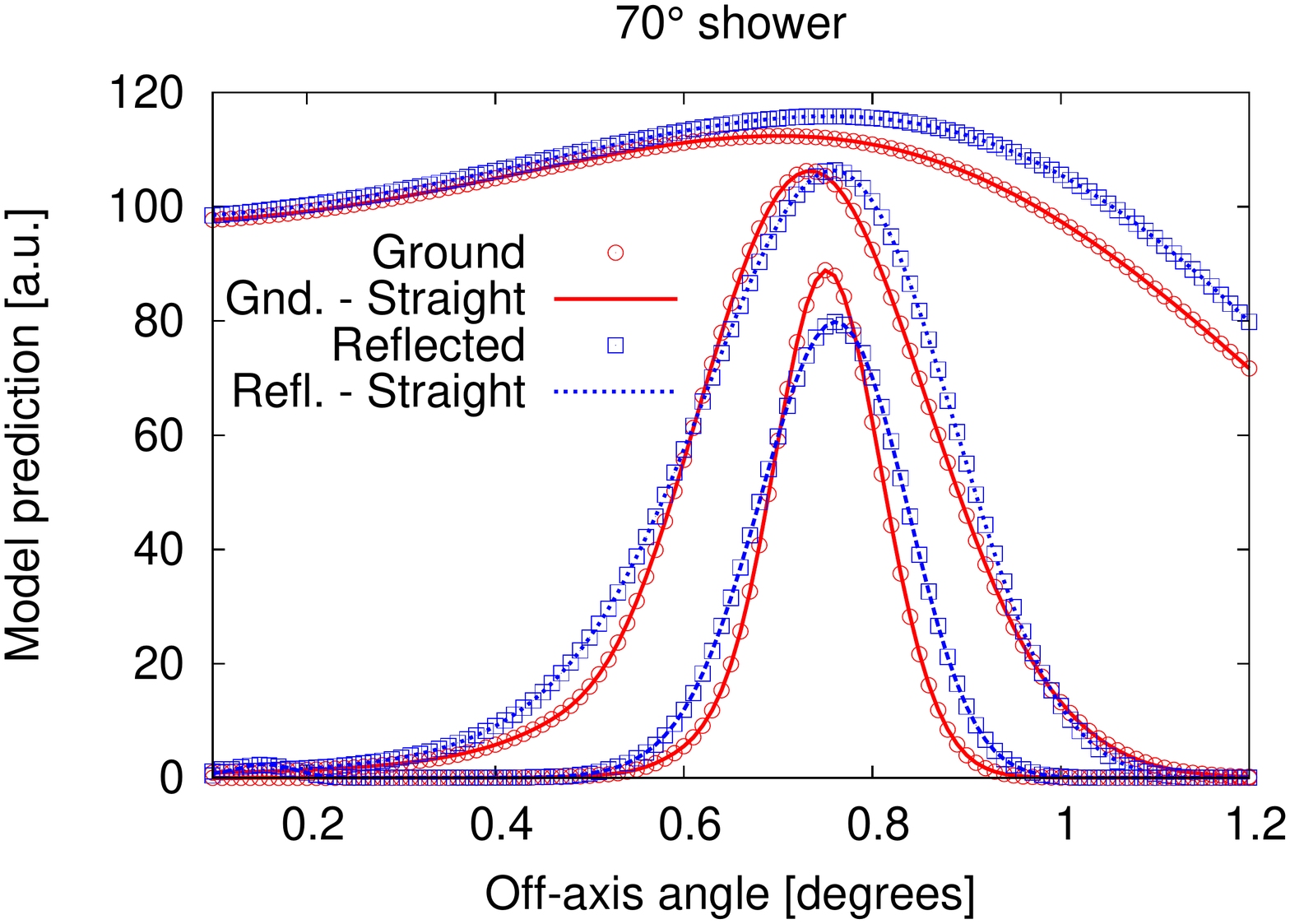} \\
\includegraphics[width=0.7\textwidth]{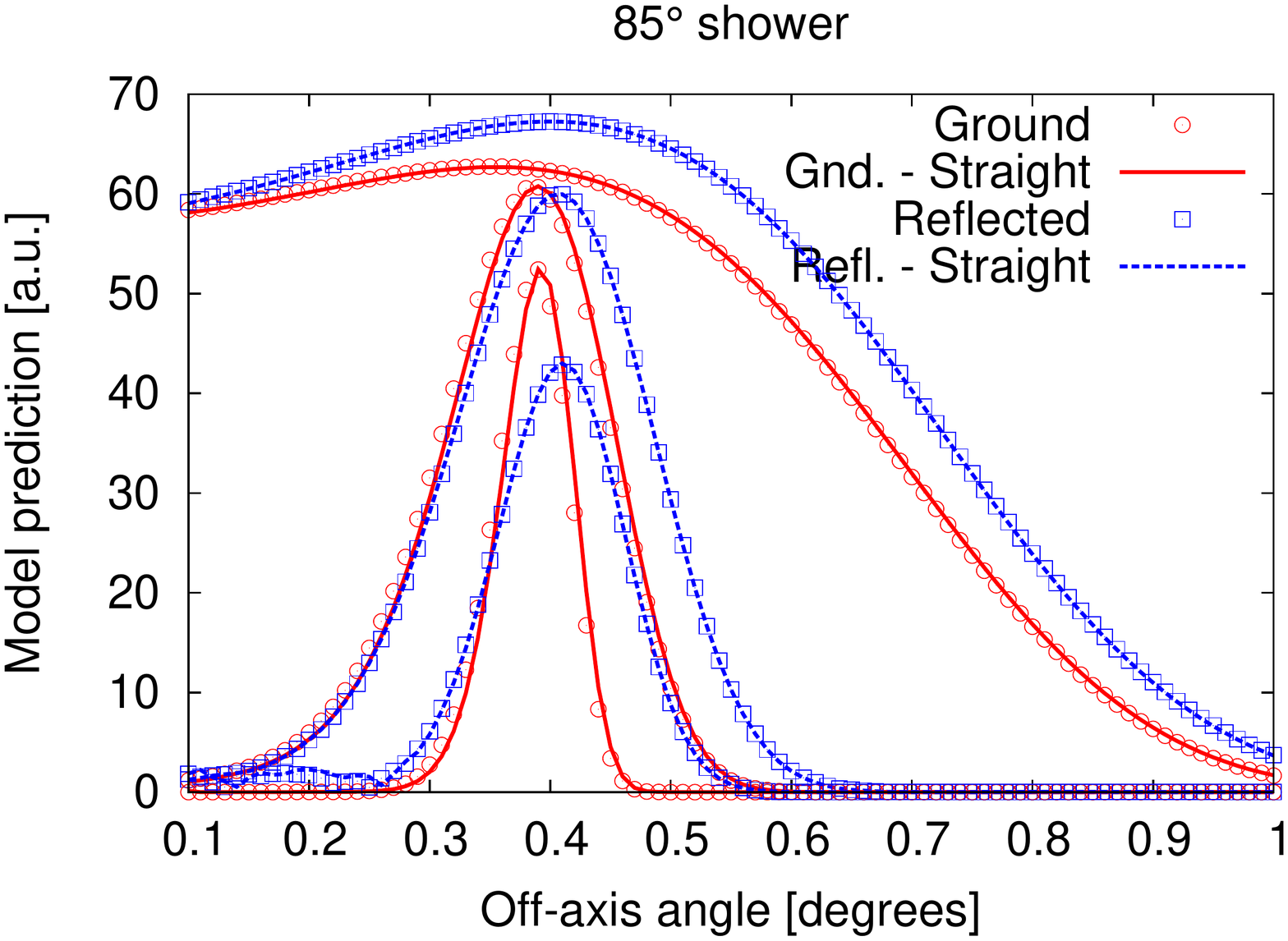}
}
\caption{Top: Electric field modulus as obtained with the simple model 
in Eq.~(\ref{eq:minimalistic_field}), as a function of the off-axis angle $\psi$
for three frequencies. The observers are located at different $\psi$ angles 
on the ground (red solid lines) and at an overall distance $\sim 130~{\rm km}$ to $X_{\rm max}$ 
after reflection (blue dashed lines). 
The shower has $\theta=70^\circ$ and $X_{\rm max}$ at an altitude above ground $h_{\rm Xmax}\sim 10.4~{\rm km}$.
The results of curved (points) and straight (lines) rays calculation are shown. 
From top to bottom, the observation frequencies are $50$, $300$ and $900~{\rm MHz}$. 
Bottom: the same as in the top panel with the observers located on the ground and at an
overall path distance of $\sim 648.5~{\rm km}$. The shower has $\theta=85^\circ$ and $h_{\rm Xmax}\sim 16.5~{\rm km}$.
In both panels the fields on the ground are rescaled with the corresponding ratio of the distance to the ground
and the total distance to the detector for visibility.} 
\label{fig:check_70deg}
\end{figure}

The results using the straight and curved
ray calculations are compared in Fig.~\ref{fig:check_70deg} where we plot 
the modulus of the electric field as a function of the offset angle of the antenna $\psi$ 
for showers of $\theta=70^\circ$ (top) and $\theta=85^\circ$ (bottom).
As anticipated from arguments concerning the travel times discussed at
the beginning of this Appendix, the difference in the angular distribution
of the electric field between the straight and curved
ray propagation is negligible for the $\theta=70^\circ$ shower
at all the frequencies we probed. It can be
appreciated that even for $\theta=85^\circ$, the effect is still negligible 
up to a frequency of $900~{\rm MHz}$. 
This justifies the straight ray approximation 
for the calculation of the angular distribution of the field,
even at high zenith angles and up to GHz frequencies.

\begin{thebibliography}{00}
%
\bibitem{CODALEMA} D. Ardouin {\it et al.} [CODALEMA Collaboration], Astropart. Phys. {\bf 31} (2009) 192.
%
\bibitem{LOFAR}
P. Schellart {\it et al.} [LOFAR Collaboration],
Astron. \& Astrophys. {\bf 560} (2013) A98.
%
\bibitem{AERA}
F.G. Schr\"oder for the Pierre Auger Collaboration,
Proceedings of the $33^{\rm rd}$ ICRC, Rio de Janeiro, Brazil (2013), \#0899.
%
\bibitem{LOPES}
H. Falcke {\it et al.} [LOPES Collaboration],
Nature {\bf 435} (2005) 313.
%
\bibitem{Tunka-Rex_ICRC13}
F.G. Schr\"oder for the Tunka-Rex Collaboration,
Proceedings of the $33^{\rm rd}$ ICRC, Rio de Janeiro, Brazil (2013), \#0452.
%
\bibitem{Huege_ICRC13}
T. Huege, Proceedings of the $33^{\rm rd}$ ICRC, Rio de Janeiro, Brazil (2013),
Braz. J. Phys. {\bf 44} (2014) 520, also available as arXiv:1310.6927 [astro-ph].
%
\bibitem{CROME}
R. {\v S}m\'\i da {\it et al.} [CROME collaboration],
Phys. Rev. Lett. {\bf 113} (2014) 221101
%
\bibitem{ZHAireS_ANITA}
J. Alvarez-Mu\~niz, W.R. Carvalho, A. Romero-Wolf, M. Tueros and E. Zas,
Phys. Rev. D {\bf 86} (2012) 123007
%
\bibitem{ANITA_CR} 
S. Hoover {\it et al.}, Phys. Rev. Lett. {\bf 105} (2010) 151001.
%
\bibitem{GeoEffect} 
F.D~Kahn, I.~Lerche, Proc. R. Soc. Lond. A {\bf 289}, 1417 (1966) 206
%
\bibitem{ZHS91} F. Halzen, E. Zas, T. Stanev, Phys. Lett. B {\bf 257} (1991) 432.
%
\bibitem{ZHS_nus} J. Alvarez-Mu\~niz, R.A. V\'azquez and E. Zas, 
Phys. Rev. D {\bf 61} (1999) 023001.
%
\bibitem{ref:Motloch}
P. Motloch, N. Hollon, P. Privitera, Astropart. Phys. {\bf 54} (2014) 40. 
%
\bibitem{EVA} 
P.~W.~Gorham, {\it et al.} 
Astropart. Phys. {\bf 35} (2011) 242. 
%
\bibitem{ref:SWORD} 
A. Romero-Wolf {\it et al}, arXiv:1302.1263 [astro-ph].
%
\bibitem{ZHAireS} J. Alvarez-Mu\~niz, W.R. Carvalho and E. Zas, Astropart. Phys., 35, 325, (2012).
%
\bibitem{1Dmodel} J. Alvarez-Mu\~niz, R.A. V\'azquez and E. Zas, Phys. Rev. D {\bf 62} (2000) 063001.
%
\bibitem{AugerXmax} P. Abreu {\it et al.} [The Pierre Auger Collaboration], Phys. Rev.  Lett. {\bf 104} (2010) 091101.
%
\bibitem{AugerXmax2} A. Aab {\it et al.} [The Pierre Auger Collaboration], arXiv1409.4809 [astro-ph].
%
\bibitem{ref:Aires} S. J. Sciutto, arXiv 99.11.331, (1999), http://www2.fisica.unlp.edu.ar/auger/aires. 
%
\bibitem{ref:Zas1992} E. Zas, F. Halzen, T. Stanev, Phys. Rev. D {\bf 45}, 362 (1992).
%
\bibitem{ref:Muniz2010} J. Alvarez-Mu\~niz, A. Romero-Wolf, E. Zas, Phys. Rev. D {\bf 81} (2010) 123009.
%
\bibitem{DanielZHS} D. Garc\'\i a-Fern\'andez, J. Alvarez-Mu\~niz,
  W.R. Carvalho, A. Romero-Wolf and E. Zas, Phys. Rev. D {\bf 87} (2013) 023003.
%
\bibitem{ZHS_time}
J. Alvarez-Mu\~niz, A. Romero-Wolf, E. Zas,
Phys. Rev. D {\bf 81} (2010) 123009.
%
\bibitem{JupiterProbe}
A. Romero-Wolf, S. Vance, F. Maiwald, E. Heggy, P. Ries, K. Liewer, arXiv:1404.1876 [astro-ph]. 
%
\bibitem{Besson_ref_index}
I. Kravchenko,  D. Besson, J. Meyers, 
Journal of Glaciology, {\bf 50} (2004) 171.
%
\bibitem{Stockham} 
D.~Z.~Besson, {\it et al.}
Radio Science (in press) arXiv:1301.4423 [astro-ph].
%
\bibitem{BelovARENA}
K. Belov for the ANITA collaboration, AIP Conf. Proc. {\bf 1535} (2013) 209. 
%
\bibitem{Washington_ARENA12}
J. Alvarez-Mu\~niz, W.R. Carvalho Jr., A. Romero-Wolf, M. Tueros, E. Zas,
AIP Conf. Proc. {\bf 1535} (2013) 143.
%
\bibitem{Scholten_MGMR}
O. Scholten, K. Werner, F. Rusdyi, Astropart. Phys.  {\bf 29} (2008) 94.
%
\bibitem{werner08} K. Werner, O. Scholten, Astropart. Phys. {\bf 29} (2008) 393.
%
\end{thebibliography}
\end{document}